%% file: main.tex
\documentclass[twocolumn]{aastex631}
\usepackage{natbib}
\usepackage{graphicx}
\usepackage{graphics}
\usepackage{tabularx}
\usepackage{lipsum}
\usepackage{subfigure}
\usepackage{amsmath,amsfonts,amsthm,bm} 
\usepackage{float}
\usepackage{bm}
\usepackage{courier}
\usepackage{color}
\usepackage{colortbl}
\usepackage{threeparttable}
\usepackage{hyperref}
\hypersetup{colorlinks=true,
linkcolor=red,
citecolor=blue}
\usepackage{blindtext}
\usepackage{placeins}
\usepackage[super]{nth}
\usepackage{multirow}
\usepackage{enumitem}
\usepackage{xcolor}

\singlespace

\newcommand{\mbf}[1] {$\mathbf #1$}

\begin{document}

\title{The Atacama Cosmology Telescope: Semi-Analytic Covariance Matrices for the DR6 CMB Power Spectra}

\input{authors}

\begin{abstract}
The Atacama Cosmology Telescope Data Release 6 (ACT DR6) power spectrum is expected to provide state-of-the-art cosmological constraints, with an associated need for precise error modeling. In this paper we design, and evaluate the performance of, an analytic covariance matrix prescription for the DR6 power spectrum that sufficiently accounts for the complicated ACT map properties. We use recent advances in the literature to handle sharp features in the signal and noise power spectra, and account for the effect of map-level anisotropies on the covariance matrix. In including inhomogeneous survey depth information, the resulting covariance matrix prescription is structurally similar to that used in the \textit{Planck} Cosmic Microwave Background (CMB) analysis. We quantify the performance of our prescription using comparisons to Monte Carlo simulations, finding better than $3\%$ agreement. This represents an improvement from a simpler, pre-existing prescription, which differs from simulations by $\sim16\%$. We develop a new method to correct the analytic covariance matrix using simulations, after which both prescriptions achieve better than $1\%$ agreement. This correction method outperforms a commonly used alternative, where the analytic correlation matrix is assumed to be accurate when correcting the covariance. Beyond its use for ACT, this framework should be applicable for future high resolution CMB experiments including the Simons Observatory (SO).
\end{abstract}

\section{Introduction} \label{sec: intro}

The Atacama Cosmology Telescope's (ACT) \citep{ACT_telescope, Thornton2016, AdvACT} sixth data release (DR6) promises to provide competitive constraints on cosmological models. The measured primary CMB anisotropy power spectrum will be used to test the $\Lambda$CDM model and to look for evidence for beyond-$\Lambda$CDM physics, including for example an early dark energy (EDE) component \citep{Hill2022} or self-interacting neutrinos \citep{Kreisch2024}. The characterization of the power spectrum covariance matrix is critical to these efforts. The pseudo-$C_{\ell}$ Monte Carlo Apodized Spherical Transform Estimator (MASTER) \citep{MASTER} formalism provides a nearly optimal, unbiased method for power spectrum reconstruction \citep{Efstathiou2004}. A key feature is its ability to correct for a common real-world systematic effect --- the incomplete sky coverage of surveys --- via the computation of ``mode-coupling" matrices. Moreover, it is computationally tractable: the computation of these matrices scales with survey resolution as $\mathcal{O}(\ell_{max}^3)$ \citep{Efstathiou2004, Toeplitz}, the same scaling as the spherical harmonic transform (SHT) \citep{libsharp}. For these reasons, MASTER has proven versatile, providing the basis for an array of results from CMB analyses \citep[e.g.,][]{Larson2011, Planck_V_2018, C20, Efstathiou2021, Dutcher2021, Balkenhol2023}, as well as clustering of galaxies and quasars, and weak lensing \citep[e.g.,][]{DESY3_2022, HSCY3_2023, Alonso2023, Piccirilli2024,DESY3_2024}. The ACT DR6 power spectrum pipeline uses the MASTER formalism.

It has long been recognized that a direct calculation of the MASTER covariance is intractable \citep[see e.g.,][]{Efstathiou2004, Challinor2004, Brown2005}. In particular, incomplete survey coverage couples to spatial correlations in the underlying field of interest; a fast approximation (that is, also $\mathcal{O}(\ell_{max}^3)$ complexity) is only possible assuming the coverage is sufficiently uniform \citep[see e.g.,][]{Couchot2017, Garcia2019}. Analyses utilizing the MASTER formalism have thus investigated the validity of this assumption for their specific application \citep[e.g.,][]{Planck_XI_2015, DESY3_2021}. The community continues to develop approaches to handle non-uniform survey geometry more robustly \citep[e.g.,][]{INKA, Camphuis2022}. The ACT DR6 data present a new challenge: the atmosphere induces stronger noise correlations than are present in space-based microwave observatories such as \textit{Planck} or galaxy surveys with close-to-Poisson noise. The data are not adequately described by the simple model required by the MASTER covariance formalism.

In this paper, we investigate the performance of a MASTER-compatible covariance matrix for the ACT DR6 power spectrum. The covariance structure resembles that of \citet{Planck_V_2018} in including inhomogeneous survey depth, and uses the analytical framework presented in \citet{INKA}. We also develop a new method for handling anisotropies introduced in the Fourier-space filtering of the ACT data. We test this covariance matrix alongside a commonly-used analytic version that assumes homogeneous survey depth, and find that the former performs better as measured by agreement with a Monte Carlo covariance matrix. Lastly, we develop a new procedure using the Monte Carlo covariance to apply a smooth correction to either matrix version and achieve sub-percent agreement with simulations.

The outline of this paper is as follows: we introduce the ACT DR6 data, and the power spectrum covariance matrix products, in \S\ref{sec: data}. In \S\ref{sec: formalism}, we discuss the MASTER framework, focusing on the particular complications of the ACT data. An overview of our covariance matrix pipeline is provided in \S\ref{sec: pipeline}, and we assess the results of that pipeline in \S\ref{sec: results}. We conclude in \S\ref{sec: conclusion}.

\section{Data and Deliverables} \label{sec: data}

\begin{table}
    \centering
    \movetableright=-42pt
    \begin{tabular}{c|c|c|c}
        Array & Band & Frequencies & Beam \\
        & & (GHz) & (arcmin) \\
        \hline
        \hline
        \multirow{2}{*}{PA4} & f150 & 124 -- 172 & 1.4 \\
        & f220 & 182 -- 277 & 1.0 \\
        \multirow{2}{*}{PA5} & f090 & 77 -- 112 & 2.0 \\
        & f150 & 124 -- 172 & 1.4 \\
        \multirow{2}{*}{PA6} & f090 & 77 -- 112 & 2.0 \\
        & f150 & 124 -- 172 & 1.4 \\
    \end{tabular}
    \caption{Frequency coverage (the 0.5\% and 99.5\% locations of their cumulative bandpower) and resolution (beam full-width half-maximum in arcminutes) of the Advanced ACT detector arrays used in DR6. Table from \citet{mnms}.}
    \label{tab: arrays}
\end{table}

Between 2007 and 2022, ACT observed the microwave sky from Cerro Toco in the Atacama Desert, Chile. ACT DR6 comprises the data collected by the Advanced ACT receiver from 2017 onwards \citep{AdvACT, PA4_yaqiong, PA4, PA56, Crowley2018}; the power spectrum analysis uses the exclusively-nighttime data (between 23:00 and 11:00 UTC) from three dichroic, polarization-sensitive detector arrays abbreviated ``PA4," ``PA5," and ``PA6." Their beam and approximate observed frequencies --- labeled by a frequency ``band" --- are given in Table \ref{tab: arrays}. For each frequency band on each array, ACT maps four disjoint ``splits" of the raw data --- formed by separately allocating each day of observations to a given split --- such that the noise in each split is independent. The maps are produced for each of the Stokes I, Q, and U polarization components in CMB blackbody temperature units. The power spectrum analysis, including this paper, uses an updated map version, ``dr6.02," which features lower noise and reduced systematics compared to the dr6.01 map version used in \citet{mnms, Qu2024, Coulton2024}.

\begin{figure*}
    \centering
    \includegraphics[width=0.85\textwidth]{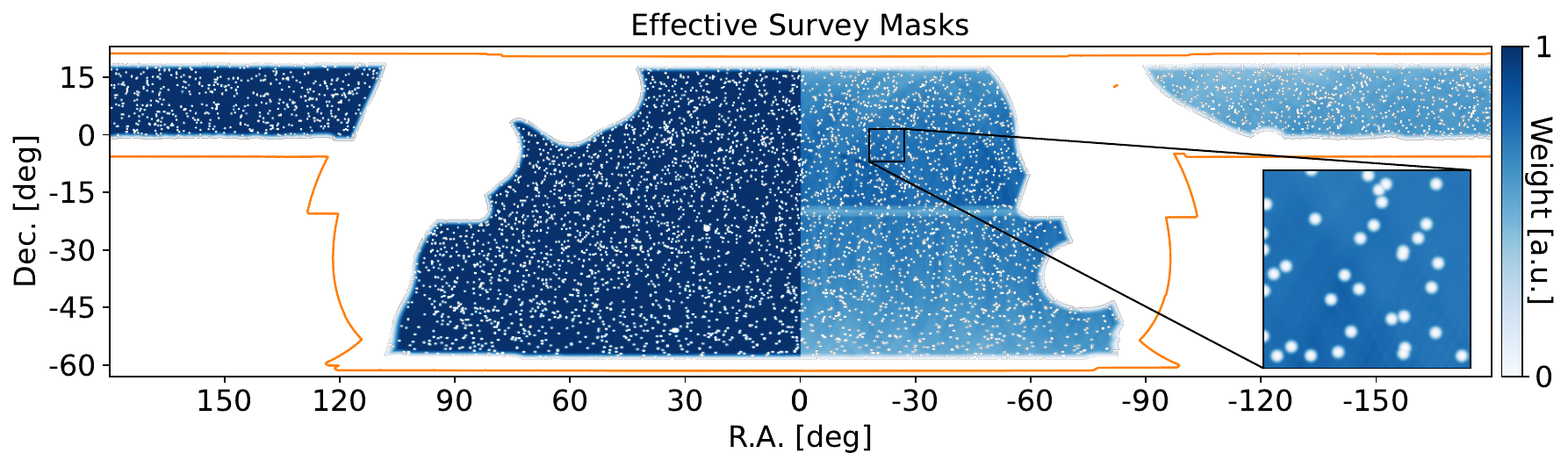}
    \caption{\textit{Blue, Left:} The power spectrum pipeline analysis mask for PA6 f150. \textit{Blue, Right:} The same mask after including effective noise weights (arbitrarily normalized), described in \S\ref{sec: pipeline_analytic}. The inset provides a zoomed-in view of the point-source holes. The outer mask borders (the point-source holes) have a $2^\circ$ ($0.3^\circ$) cosine apodization. \textit{Orange:} The outline of the ACT survey footprint. Data within the orange outline, but not highlighted in blue, are excluded from the analysis.}
    \label{fig: data_masks}
\end{figure*}

Accompanying the data maps are ``inverse-variance maps" \citep[see e.g.][]{A20, MK21}, which give the white-noise level of the data. In \S\ref{sec: pipeline}, we use these to derive maps of the white-noise standard deviation per pixel by taking their inverse square-root. The DR6 power spectrum pipeline also includes analysis masks which are applied to the maps, with an example shown in Figure \ref{fig: data_masks}. The masks have apodized boundaries to limit mode-coupling, and avoid both the edge of the ACT survey footprint where data is noisiest as well as the plane of the Galaxy where foregrounds are brightest. The masks also contain $\sim10,000$ apodized point-source holes with a diameter of five arcmin. These correspond to a catalog of sources with measured flux greater than 15 mJy in the 150\,GHz channel and detected with a significance greater than $5\sigma$.

As will be described in forthcoming papers, the ACT DR6 power spectrum analysis includes substantial data validation that took place with blinded results, resulting in the removal of PA4 f150 (temperature and polarization) and PA4 f220 polarization from the analysis. Scale cuts are applied to the remaining data that range from a minimum angular multipole of 475 -- 975, depending on array and polarization. The maximum multipole is $\ell_{max}=8,500$ for all arrays. The spectra are binned following the same schedule as presented in \citet{C20}, Appendix F. Each polarized cross spectrum contains up to 57 bins, and the entire data vector including all $T$, $E$, and $B$ spectra, has over 3,500 elements. The baseline DR6 likelihood only uses $TT$, $TE$, and $EE$ spectra; after the cuts we use in this paper, the data vector has 1,763 elements.\footnote{Minor revisions were made to the dr6.02 maps, analysis masks, scale cuts, fiducial signal spectra (\S\ref{sec: pipeline_analytic_signal}), and simulations (\S\ref{sec: pipeline_sims}) after the completion of this paper, but should not affect our conclusions.} 

A ``first version" covariance matrix for the ACT power spectrum was constructed using pre-existing tools available in \texttt{pspy}\footnote{\url{https://github.com/simonsobs/pspy}} and \texttt{PSpipe}.\footnote{\url{https://github.com/simonsobs/PSpipe}} We label this matrix the ``homogeneous" matrix for reasons discussed in \S\ref{sec: pipeline}. In this paper we develop a new covariance matrix; we label it the ``inhomogeneous" matrix. Both versions start with an ``analytic" covariance matrix, which is then corrected using simulations. For clarity, we refer to the post-correction matrix as the ``corrected" or ``semi-analytic" matrix. This semi-analytic matrix is the product used in the likelihood to constrain cosmological models. The homogeneous matrix is the default version used in the baseline DR6 likelihood, as it was developed earlier, but as we show in this paper, we find that our new inhomogeneous version requires a smaller simulation-based correction --- in other words, that its analytic prescription better describes the data covariance. 

\section{Analytical Framework} \label{sec: formalism}

In this section we describe a model for the ACT DR6 data, as well as approximations  that simplify the construction of the power spectrum covariance matrix.

\subsection{ACT Data Model} \label{sec: formalism_full_data_model}

The ACT data contain several properties that, in principle, are challenging to incorporate into the MASTER framework. We start by describing a generative model for the ACT data units: the 20 polarized split maps enumerated in \S\ref{sec: data}. Most ACT map metadata are bundled into an ``array" label, but for the formalism it is useful to keep the map polarization and split index separate. Thus, we label a given map, \mbf{m}, as \mbf{m_{I_i}^{X}}, where $I$ denotes the array, $i$ the split index, and $X$ the polarization component. We consider the map to be a sum of signal and noise components: 
\begin{align} \label{eq: split_def}
    \mathbf m_{I_i}^{X} = \mathbf s_I^{X} + \mathbf n_{I_i}^{X}
\end{align}
where the signal does not depend on the map split.\footnote{This is true insofar as the beam and passband are the same for each split of an array. Beam variations over split in dr6.02 maps are $\mathcal{O}(0.1\%)$ at signal-dominated scales and so we neglect them. There is no evidence for split-dependent passband variations.}

We assume the sky signal to be a realization from a homogeneous, isotropic Gaussian distribution that is subsequently modified by the instrument.\footnote{When discussing a field throughout the paper, we take ``inhomogeneous" to mean ``spatially-dependent variance over 2D positions in a map" and ``anisotropic" to mean ``azimuthally-dependent covariance when at a fixed 2D position in a map."} Its power spectrum includes contributions from the primary CMB, as well as secondary effects including CMB lensing, the thermal and kinetic Sunyaev Zeldovich effects (tSZ and kSZ), the cosmic infrared background (CIB), unresolved (``point source") active galactic nuclei (AGN), and dust. The sky signal is passed through the telescope, receiver optics, and filtering, and is sampled by the detector arrays at the focal plane. This process integrates the frequency dependence of each component over each array’s spectral passband and spatially convolves the continuous sky with the instrumental beam.\footnote{To streamline \S\ref{sec: formalism}, we do not discuss other effects in the data, such as calibration and pixel window functions, here. Those are detailed in Appendix \ref{apx: sims}.} We assume transient instrumental and astrophysical sources can be modeled by their time-averaged quantities over the course of ACT observations. 

\begin{figure*}
    \centering
    \includegraphics[width=0.85\textwidth]{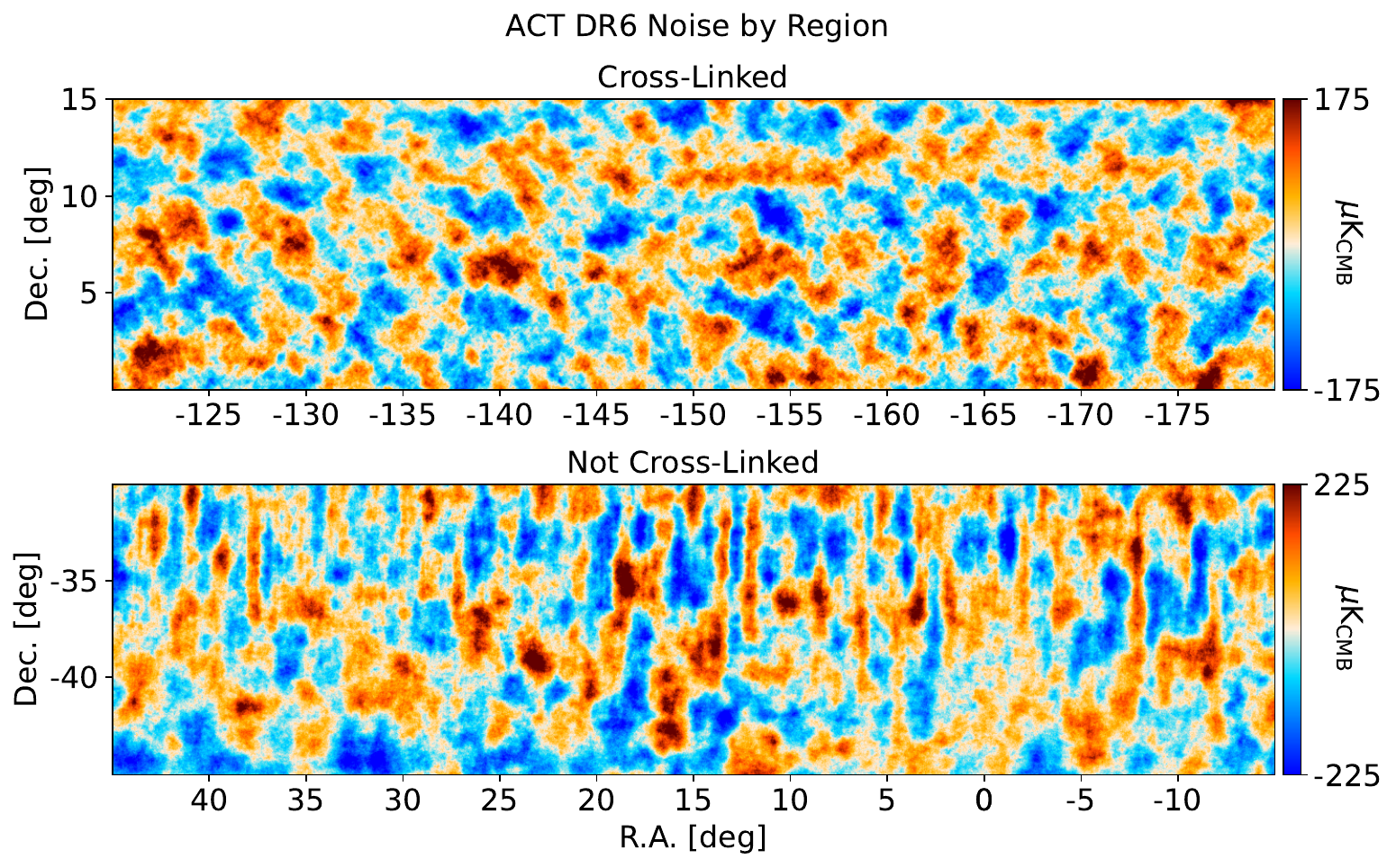}
    \includegraphics[width=0.85\textwidth]{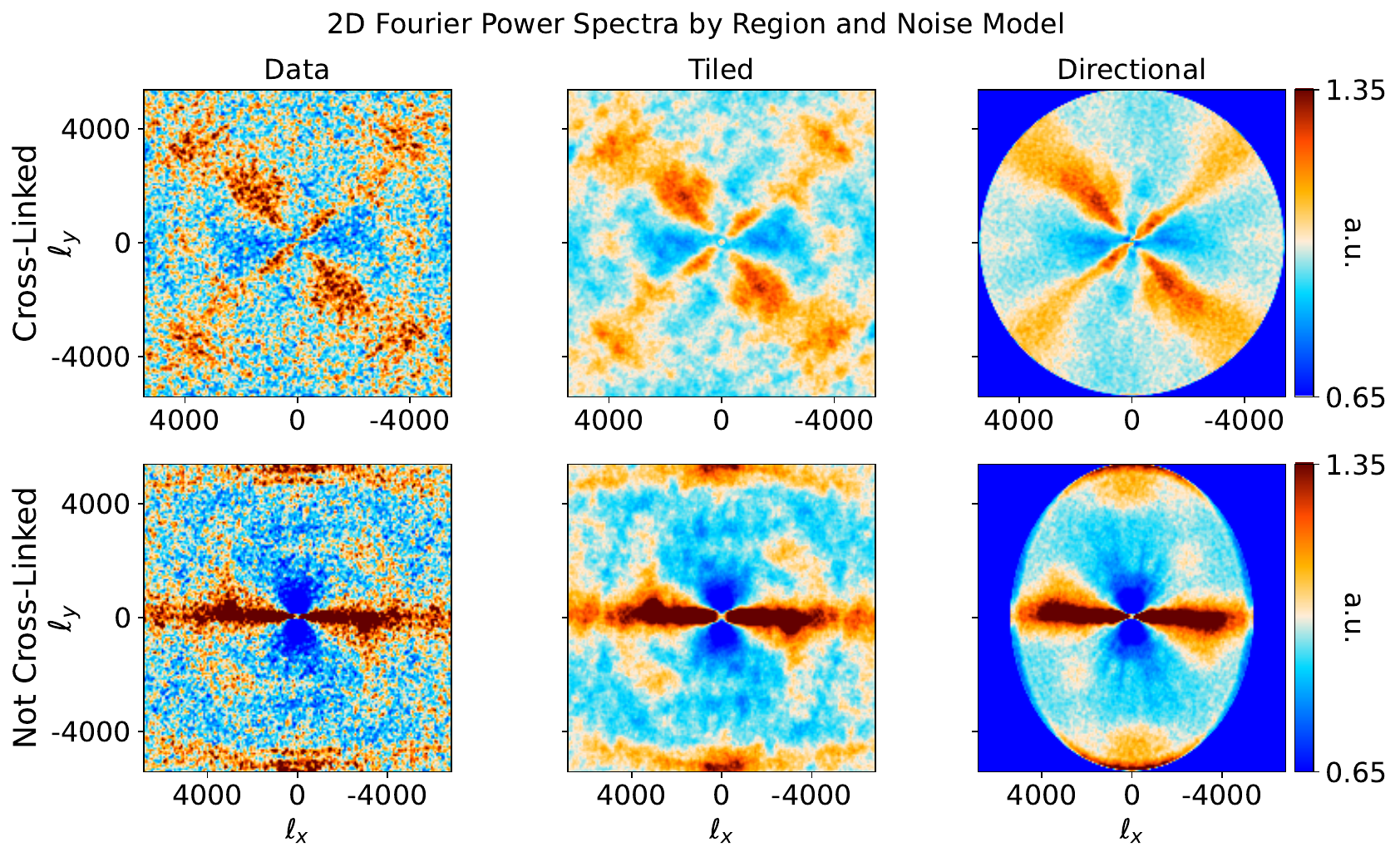}
    \caption{\textit{First and second rows:} The noise in the first temperature split map for PA5 f090 measured in a $900$ deg$^2$ well-cross-linked region of the ACT scan strategy. The second row shows a region with less cross-linking where scans only move in the vertical (Dec.-only) direction. \textit{Third and fourth rows, left:} 2D Fourier noise power spectra of the first temperature split map for PA5 f090. The average radial profiles of the power spectra have been divided-out to better highlight their anisotropic patterns. The third row is measured in the well-cross-linked region, where crossing scans induce the ``x-like" bars in the power spectrum. The fourth row is measured in the poorly-cross-linked region, where the Dec.-aligned scans induce vertical noise stripes in the maps that appear as a horizontal bar in the power spectrum. \textit{Center and right:} 2D power spectra from noise simulations following \citet{mnms} (drawn from the tiled and directional wavelet models). The tiled model has a bandlimit of $\ell_{max}=10,800$; the wavelet model has $\ell_{\max}=5,400$, visible as a hard edge in 2D Fourier space.}
    \label{fig: form_pa5_f090_set0_2d_T_sims}
\end{figure*}

Collectively, we have the following model of the signal:
\begin{align} \label{eq: signal_def}
    \mathbf s = \mathbf Y \mathbf B \mathbf{S}^{\frac{1}{2}} \bm\eta_s,
\end{align}
where we have suppressed the array ($I$) and polarization ($X$) labels for simplicity. We read from right to left, starting with $\bm\eta_s$, which is a white-noise vector in harmonic space. Thereafter, each symbol is a linear operator that acts on the vector: $\mathbf{S}$ is the power spectrum of the sky signal with elements that covary over arrays (according to their passbands) and polarizations,\footnote{The matrix exponent ensures that $\mathbf{S}^{\frac{1}{2}} \bm\eta_s$ is a \textit{sample} of spherical harmonic coefficients whose covariance is the power spectrum.} \mbf{B} is the isotropic beam transfer function\footnote{We do not measure significant beam anisotropy in DR6 and do not account for any in this paper.} for each array and polarization, and \mbf{Y} is a spherical harmonic transform (SHT) ``synthesis" operation (i.e., the matrix that transforms vectors from harmonic space to map space). Note that both $\mathbf{S}^{\frac{1}{2}}$ and \mbf{B} are assumed to be diagonal over spherical harmonic modes, such that this model of the map signal is, like the underlying sky, homogeneous, isotropic, and Gaussian.

Like the signal, the map-level noise is Gaussian distributed, but is inhomogeneous and anisotropic: the noise power varies over the sky, and its correlation structure is ``stripy." Moreover, the noise anisotropy is itself inhomogeneous: the stripy correlation pattern varies over the sky (see \S3 of \citet{mnms} for a complete description, and Figure \ref{fig: form_pa5_f090_set0_2d_T_sims} for illustration). These properties are the product of projecting correlated atmospheric noise along the ACT scanning direction.

We model the noise at the map level following \citet{mnms}, which describes two models that capture the most relevant features: a ``tiled" and a ``directional wavelet" noise model. The tiled model subdivides the map into a set of interleaved patches (``tiles") and builds the noise covariance in each tile's 2D Fourier space. The wavelet model instead subdivides the Fourier transform of the map into interleaved patches (``wavelets") and builds the noise covariance in each wavelet's map space. These noise models are not necessarily complete, but simulations drawn from them should be sufficiently accurate for correcting and evaluating the covariance matrix. Figure \ref{fig: form_pa5_f090_set0_2d_T_sims} demonstrates an example of the efficacy of each model at reproducing the spatially-varying noise stripiness in the data. We are agnostic as to which model performs best; rather, their existence allows us to assess robustness of the covariance matrix to the assumed noise model. Time-domain-based simulations, a possible alternative, are too computationally expensive to be practical.

As described in \cite{mnms} and Appendix \ref{apx: mnms}, samples drawn from either model follow similar prescriptions:
\begin{equation}  \label{eq: noise_def}
    \begin{aligned}
        \mathbf n_i &= \bm\sigma_i \mathbf Y \mathbf N_i^{\frac{1}{2}} \mathbf Y^{\dag}\bm\Omega (\mathcal{T}_b^{\dag} \mathbf F^{\dag} \mathcal{N}_i^{\frac{1}{2}})\bm\eta_{n,i} \quad \text{(Tiled)} \\
        \mathbf n_i &= \bm\sigma_i \mathbf Y \mathbf N_i^{\frac{1}{2}} \mathbf Y^{\dag}\bm\Omega (\mathbf F^{\dag} \mathcal{T}_b^{\dag} \mathbf F \mathcal{N}_i^{\frac{1}{2}})\bm\eta_{n,i}, \quad \text{(Dir. Wavelet)}
    \end{aligned}
\end{equation}
where, again, we have suppressed the array ($I$) and polarization ($X$) labels for simplicity. Reading from right to left, $\bm\eta_n$ is a white-noise vector, $\mathcal{N}$ is the noise covariance in either tiled 2D Fourier space or wavelet map space, \mbf{F} is the unitary Discrete Fourier Transform (DFT), and $\mathcal{T}_b$ is the ``backward tiling transform" that reverts the map or Fourier space subdivision (for the tiled model or directional wavelet model, respectively; see \citet{mnms} for more technical detail). The terms in parentheses are the defining feature of either noise model: they are what introduce the spatially-varying noise anisotropy. The sample then undergoes spherical harmonic ``analysis" (i.e., from map to harmonic space), denoted by $\mathbf Y^{\dag}\bm\Omega$, where $\bm\Omega$ is a diagonal matrix in map space containing spherical harmonics analysis quadrature weights that approximately equal the area of each pixel. Finally, \mbf{N} is the estimated noise power spectrum, and $\bm\sigma$ contains the per-pixel noise standard deviations from \S\ref{sec: data}.\footnote{The noise power spectrum, \mbf{N}, is estimated after normalizing the maps by $\bm\sigma$.} The noise realization, \mbf{n}, as well as components of the noise model, depend on the map split $i$. We emphasize that a realistic model of the ACT map noise, compared to the signal model in Equation \ref{eq: signal_def}, is inhomogeneous and anisotropic.

Two steps in the ACT DR6 analysis that affect the data model are the application of a Fourier-space filter and an apodized analysis mask to the data maps. Fourier modes are removed from the maps for which $|\ell_x|\leq 90$ and $|\ell_y|\leq 50$; this is the same filter as in ACT DR3 \citep{L17} and DR4 \citep{C20}. This filter removes pickup at large scales in the horizontal (R.A.-aligned) direction, and noise-dominated modes at large scales in the vertical (Dec.-aligned) direction. This filter introduces anisotropy to the signal component of the maps, while adding to the preexisting map noise anisotropy. The analysis masks described in \S\ref{sec: data} are applied after the Fourier-space filter.

Combining these processes with Equations \ref{eq: split_def}, \ref{eq: signal_def}, and \ref{eq: noise_def} yields the following data model for the masked maps (e.g., in the case of the ``tiled" noise model):
\begin{equation} \label{eq: full_data_model}
    \begin{aligned}
        \tilde{\mathbf m}_i &\equiv \mathbf W \mathbf F^{\dag} \mathbf X_f \mathbf F \mathbf m_i = \mathbf W \mathbf F^{\dag} \mathbf X_f \mathbf F \mathbf Y \mathbf B \mathbf{S}^{\frac{1}{2}} \bm\eta_s \\
        &+ \mathbf W \mathbf F^{\dag} \mathbf X_f \mathbf F \bm\sigma_i \mathbf Y \mathbf N_i^{\frac{1}{2}} \mathbf Y^{\dag}\bm\Omega \mathcal{T}^{\dag} \mathbf F^{\dag} \mathcal{N}_i^{\frac{1}{2}}\bm\eta_{n,i},
    \end{aligned}
\end{equation}
where \mbf{X_f} denotes the Fourier-space filter (which is diagonal in Fourier space), and \mbf{W} denotes the analysis mask (which is diagonal in map space). While realistic, this anisotropic data model is not amenable to the MASTER framework.

\subsection{Covariance Matrices in MASTER} \label{sec: formalism_MASTER}

The MASTER method includes the computation of both a power spectrum estimator, $\hat{C}_{\ell}$, and a pseudospectrum covariance matrix, $\tilde{\Sigma}_{\ell\ell'}$. For a data model consisting of an isotropic, scalar Gaussian field, $a$, sampled from the power spectrum $C_{\ell}$, and masked by an analysis window $w$ at sky positions $x$ with
\begin{align} \label{eq: scalar_MASTER_model}
    \tilde a(x) &\equiv w(x)a(x),
\end{align}
the MASTER power spectrum estimator is given\footnote{This paper, and the standard MASTER framework, assume that the window $w$ is not correlated with underlying field. Recent work by \citet{Surrao2023} has derived modifications to the framework when this assumption is invalid.} by:
\begin{equation} \label{eq: 2pt_MASTER}
    \begin{aligned}
        \hat{C}_{\ell} &= \sum_{\ell_1}M_{\ell\ell_1}^{-1}\hat{\tilde{C}}_{\ell_1}\\
        M_{\ell\ell_1} &\equiv (2\ell_1+1)\Xi_{\ell\ell_1}(w,w)\\
        \hat{\tilde{C}}_{\ell} &\equiv \frac{1}{2\ell+1} \sum_{m=-\ell}^{\ell} \tilde a_{\ell m} \tilde a_{\ell m}^*,
    \end{aligned}
\end{equation}
\citep[e.g.,][]{Efstathiou2004, Couchot2017, Garcia2019}, where $M_{\ell\ell_1}$ is the mode-coupling matrix, $\Xi_{\ell\ell_1}$ is the symmetric coupling matrix \citep{Efstathiou2004}, which we refer to as the ``coupling," $\hat{\tilde{C}}_{\ell}$ is the \textit{pseudo}spectrum estimate, and $\tilde a_{\ell m}$ is the SHT of $\tilde a(x)$. The coupling is only a function of the analysis mask via its power spectrum \citep{MASTER}, $W_{\ell} \equiv 1/(2\ell+1)\sum_m w_{\ell m} w_{\ell m}^*$. Under the data model of Equation \ref{eq: scalar_MASTER_model}, the power spectrum estimator in Equation \ref{eq: 2pt_MASTER} is exact and calculable in $\mathcal{O}(\ell_{max}^3)$-time.

\begin{figure*}
    \centering
    \includegraphics[width=0.85\textwidth]{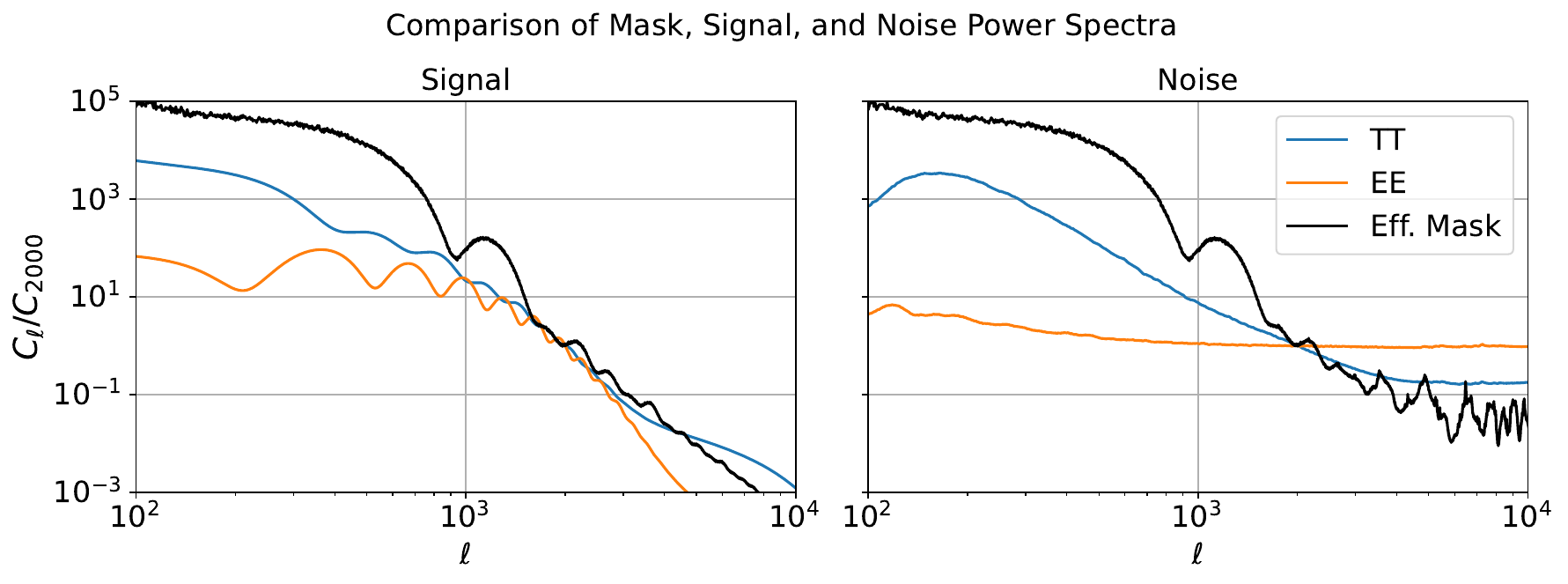}
    \caption{Signal and noise power spectra compared to the power spectra for their effective masks (for PA6 f150, first split). The difference between the signal and noise effective masks is shown in Figure \ref{fig: data_masks}: the effective mask for the noise includes the inhomogeneous survey depth and so has more structure than the signal mask, as reflected in its wider mask power spectrum. For both the signal and noise, the power spectrum of the mask does not appear to be significantly more compact, or steep, than the power spectrum of the field itself, calling the NKA into question for ACT. All power spectra are normalized at $\ell=2,000$.}
    \label{fig: form_Wl_Sl_Nl}
\end{figure*}

Unfortunately, there is no known analytical expression for the covariance of the pseudospectrum estimator that is also calculable in $\mathcal{O}(\ell_{max}^3)$-time. A naive computation leads to an expression whose complexity is $\mathcal{O}(\ell_{max}^6)$ \citep{Challinor2004}. Recently, \citet{Camphuis2022} derived an algorithm with $\mathcal{O}(\ell_{max}^5)$ complexity. Instead, many analyses \citep[see e.g.,][]{Planck_XI_2015, Planck_V_2018, DESY3_2022, DES_HSC_KIDS_2024} have employed a pseudospectrum covariance matrix with the following form:
\begin{align} \label{eq: 4pt_MASTER}
        \tilde{\Sigma}_{\ell\ell'} &\equiv \langle(\hat{\tilde{C}}_{\ell} - \langle\hat{\tilde{C}}_{\ell}\rangle)(\hat{\tilde{C}}_{\ell'} - \langle\hat{\tilde{C}}_{\ell'}\rangle)\rangle \approx 2C_{(\ell,\ell')}^2\Xi_{\ell\ell'}(w^2,w^2),
\end{align}
where the $(\ell,\ell')$ subscripts denotes some symmetric function of $C_{\ell}$ and $C_{\ell'}$. As it uses the coupling matrix from Equation \ref{eq: 2pt_MASTER}, this expression is also calculable in $\mathcal{O}(\ell_{max}^3)$-time; however, it requires invoking the ``narrow kernel approximation" \citep[NKA, see][]{Garcia2019}, which states that the power spectrum of the mask, $W_{\ell}$, is significantly more compact in $\ell$ than that of the underlying field. For the mask, signal, and noise power spectra in ACT, it is not clear whether this approximation is valid (see Figure \ref{fig: form_Wl_Sl_Nl}).

Furthermore, regardless of the NKA, the MASTER formalism requires a simple data model like that in Equation \ref{eq: scalar_MASTER_model}. Expressed in the symbolic formalism of \S\ref{sec: formalism_full_data_model}, this model looks like:
\begin{align} \label{eq: vector_MASTER_model}
    \mathbf a = \mathbf W \mathbf Y \mathbf{C}^{\frac{1}{2}} \bm\eta,
\end{align}
where \mbf{C} is a diagonal matrix whose elements are given by $C_{\ell}$. As defined in Equation \ref{eq: full_data_model}, ACT's signal field differs from this simpler model due to the Fourier-space filter, while the noise field also differs due to its intrinsically inhomogeneous and anisotropic structure. Thus, in order to work with MASTER methods, we need an approximate data model whose signal and noise better resemble Equation \ref{eq: vector_MASTER_model}, at the cost of reduced realism.

\subsection{Approximate Data Model} \label{sec: formalism_approx_data_model}

Our covariance matrix can only involve analysis masks and isotropic power spectra as inputs (as in Equation \ref{eq: vector_MASTER_model}). Here, we describe the approximations to Equation \ref{eq: full_data_model} that enable this.

Because the MASTER data model does not permit intrinsic inhomogeneity nor anisotropy, we construct an isotropic and homogeneous expression for the noise by eliminating the anisotropic operations from Equation \ref{eq: full_data_model}. Doing so for either noise model yields:
\begin{align} \label{eq: almost_approx_noise_def}
    \mathbf n_i = \bm\sigma_i \mathbf Y \mathbf N_i^{\frac{1}{2}} \mathbf Y^{\dag}\bm\Omega \bm\eta_{n,i},
\end{align}
where $\bm\eta_n$ is a white-noise vector in map space. As we show in Appendix \ref{apx: sims}, this noise model is approximately equal to the following expression:
\begin{align} \label{eq: approx_noise_def}
    \mathbf n_i = \bm\sigma_i \bm\Omega^{\frac{1}{2}} \mathbf Y \mathbf N_i^{\frac{1}{2}} \bm\eta_{n,i} \quad \text{(Approximate)}
\end{align}
where $\bm\eta_n$ is now a white-noise vector in \textit{harmonic} space. This noise model states that we draw a Gaussian realization from an isotropic noise power spectrum, and then weight each pixel by the square-root of its area and its noise standard deviation. While incurring a ``data model error," its ingredients are now identical in form to those of the signal, where $\bm\sigma_i \bm\Omega^{\frac{1}{2}}$ is an effective noise-weight mask. 

The Fourier-space filter also introduces anisotropy to both the signal and noise. Unlike the intrinsic noise anisotropies, the anisotropy due to the filter is too large to be neglected and must be accounted for analytically. The core of the approximation is treating the filter in harmonic space, rather than Fourier space. We give further detail on the exact approximation we use in \S\ref{sec: pipeline} and Appendix \ref{apx: covmat_kspace}; for the purposes of this section, we absorb the filter into the definition of the signal and noise power spectra as an isotropic transfer function, \mbf{T^\alpha}, in Equations \ref{eq: 2pt_MASTER} and \ref{eq: 4pt_MASTER}, where $\alpha$ is an exponent that helps control the function shape, and \mbf{T} is diagonal over $\ell$ (i.e., its diagonal is $t_{\ell}^{\alpha})$. While similar to \citet{C20}, the treatment of \mbf{T} in this paper does not assume that it enters those equations with fixed exponents, but rather we find these exponents are functions of other inputs, predominantly the filter itself. 

Taken together, we have the following approximate data model for the ACT DR6 maps:
\begin{align} \label{eq: approx_data_model}
    \tilde{\mathbf m}_i = \mathbf W \mathbf Y \mathbf B (\mathbf T^\alpha  \mathbf{S})^{\frac{1}{2}} \bm\eta_s + \mathbf W \bm\sigma_i \bm\Omega^{\frac{1}{2}} \mathbf Y (\mathbf T^\alpha \mathbf N_i)^{\frac{1}{2}} \bm\eta_{n,i}.
\end{align}
This simplified model is compatible with the MASTER formalism in analogy with Equation \ref{eq: vector_MASTER_model}. To be clear, we do not expect Equation \ref{eq: approx_data_model} to correctly model the anisotropic ACT data at the map level. Rather, it is an attempt to preserve as much of the data realism as possible while still permitting the use of the MASTER covariance framework.

As discussed in the next section, the form of the covariance matrix that results from Equation \ref{eq: approx_data_model} closely resembles that in \citet{Planck_XI_2015}. Unlike for \textit{Planck}, however, it requires assuming the approximate ACT data model is sufficiently accurate at the level of the power spectrum covariance matrix. We investigate the validity of both the NKA and the approximate data model in \S\ref{sec: results}.

We emphasize that the need for such an investigation is not abstract. Because Equation \ref{eq: approx_data_model} is incorrect for the ACT data, the covariance estimate derived from it in the next section is \textit{biased}. This bias is analogous to the ``noise bias" occurring in power spectrum estimates from maps with correlated noise. Unlike the noise bias for the power spectrum estimator, which can be circumvented through the use of cross-spectra \citep{noise_bias1, noise_bias2}, this ``covariance bias" is unavoidable and its size must be checked empirically.

\section{Covariance Matrix Pipeline} \label{sec: pipeline} 

In this section, we describe how we construct the inhomogeneous ACT DR6 power spectrum covariance matrix. First, we assemble the analytic part, assuming the NKA and the approximate data model in Equation \ref{eq: approx_data_model} (\S\ref{sec: pipeline_analytic}). In parallel, we construct an ensemble of simulations following the ACT DR6 data model of Equation \ref{eq: full_data_model} (\S\ref{sec: pipeline_sims}). Lastly, we use the Monte Carlo covariance matrix formed by the simulations to correct approximation-induced errors in the analytic matrix (\S\ref{sec: pipeline_sim_correction}). This matrix is only the disconnected, Gaussian part of the covariance; beam measurement uncertainty, and the subdominant non-Gaussian contributions to the covariance from CMB lensing, clusters, and point sources will be discussed in the ACT DR6 power spectrum paper. 

\subsection{Analytic Covariance Matrix} \label{sec: pipeline_analytic}

We generalize the MASTER pseudospectrum covariance in Equation \ref{eq: 4pt_MASTER} to the case of multiple fields and datasets. We write Equation \ref{eq: approx_data_model} in the form of Equation \ref{eq: scalar_MASTER_model}, including all metadata labels:
\begin{equation} \label{eq: general_data_model}
    \begin{aligned}
        m_{I_i}^{X}(x) &= w_I^{X}(x)s_I^{X}(x) + w_I^{X}(x)\sigma_{I_i}^{X}(x)\Omega^{\frac{1}{2}}(x)n_{I_i}^{X}(x) \\
        &\equiv u_I^{X}(x)s_I^{X}(x) + v_{I_i}^{X}(x)n_{I_i}^{X}(x),
    \end{aligned}   
\end{equation}
where we have defined the effective signal and noise masks ($u$ and $v$), $s_I^{X}(x)$ is the realization from the isotropic signal power spectrum including the beam ($S_{IJ,\ell}^{XY}$), and $n_{I_i}^{X}(x)$ is the realization from the isotropic noise power spectrum ($N_{I_i J_j,\ell}^{XY}$). We have assumed that the analysis masks ($w$) and signal realizations do not depend on the split ($i$). Thus, the map pseudospectra carry a pair of metadata labels, and the covariance carries two pairs.

Assuming the signal and noise are uncorrelated, we obtain for the pseudospectrum covariance matrix:
\begin{equation} \label{eq: general_pseudocov}
    \begin{aligned}
        &\tilde{\Sigma}_{I_i J_j,P_p Q_q,\ell\ell'}^{WX,YZ} = \\
        &= S_{IP,(\ell,\ell')}^{WY}S_{JQ,(\ell,\ell')}^{XZ}\Xi_{\ell\ell'}^{\beta(WY,XZ)}(u_I^W u_P^Y, u_J^X u_Q^Z) \\
        &+ S_{IP,(\ell,\ell')}^{WY}N_{J_j Q_q,(\ell,\ell')}^{XZ}\Xi_{\ell\ell'}^{\beta(WY,XZ)}(u_I^W u_P^Y, v_{J_j}^X v_{Q_q}^Z) \\
        &+ N_{I_i P_p,(\ell,\ell')}^{WY}S_{JQ,(\ell,\ell')}^{XZ}\Xi_{\ell\ell'}^{\beta(WY,XZ)}(v_{I_i}^W v_{P_p}^Y, u_J^X u_Q^Z) \\
        &+ N_{I_i P_p,(\ell,\ell')}^{WY}N_{J_j Q_q,(\ell,\ell')}^{XZ}\Xi_{\ell\ell'}^{\beta(WY,XZ)}(v_{I_i}^W v_{P_p}^Y, v_{J_j}^X v_{Q_q}^Z) \\
        &+ (Y, P_p) \leftrightarrow (Z, Q_q),
    \end{aligned}
\end{equation}
where we introduce the coupling ``spin", $\beta$, which can be one of ``00," ``0+", ``++", or ``$--$" \citep{Brown2005}. The dependence of $\beta$ on the four input polarizations is given in Appendix \ref{apx: covmat_core}. The argument of the coupling denotes the cross-power spectrum of a pair of ``masks" that are each a product of two effective signal or noise masks. The arrow indicates that the rest of the expression is formed by the interchange of the third and fourth fields. We derive this result in Appendix \ref{apx: covmat_core}, although it is also presented implicitly in \citet{Planck_XI_2015} and \citet{INKA}. The analogous expression for a single field (e.g., the ``signal-only" covariance), or for the special case that the effective signal and noise masks are all equivalent ($u=v$), is a standard result in the literature \citep[see e.g.,][and also our discussion in \S\ref{sec: pipeline_analytic_comparison}]{Brown2005, Efstathiou2006, Couchot2017, Garcia2019}. Besides the coupling spin, there is no fundamental difference between Equations \ref{eq: 4pt_MASTER} and \ref{eq: general_pseudocov} -- the complication is just in the (expansive) bookkeeping of signal, noise, array, polarization, and split.

We use a modified version of the traditional NKA --- the ``improved" NKA \citep[INKA,][]{INKA} --- in light of the steepness of the CMB and ACT noise spectra in Figure \ref{fig: form_Wl_Sl_Nl}. Here, each power spectrum in Equation \ref{eq: general_pseudocov} is replaced by a normalized pseudospectrum:
\begin{align} \label{eq: INKA}
    C_{IJ,\ell}^{XY} \rightarrow \tilde{C}_{IJ,\ell}^{XY} / w_2(w_I^X,w_J^Y),
\end{align}
where $C$ is either $S$ or $N$, $w_I^X$ and $w_J^Y$ are the effective masks for arrays (polarizations) $I$ and $J$ ($X$ and $Y$), and $w_2(a, b)=\sum_x a(x)b(x)\Omega(x)/4\pi$. The pseudospectra are related to the power spectra via the MASTER mode-coupling matrices: the expressions are a standard result in the literature \citep[e.g.,][]{WMAP2003, Alonso2019}. Generalizing Equation \ref{eq: 2pt_MASTER}, the mode-coupling matrices are a function of the cross-power spectrum of $w_I^X$ and $w_J^Y$. \citet{INKA} found the INKA improves the accuracy of analytic MASTER covariance matrices when applied to galaxy weak lensing, and argued for its superiority in all cases. It has seen increasing uptake in (for example) \citet{Hadzhiyska2021, Euclid2022, DESY3_2022, DESY3_2024}, but to our knowledge has not yet been used in a CMB analysis. 

As in \citet{INKA}, we choose to use the arithmetic mean to define the symmetric function $C_{(\ell,\ell')}\equiv (C_{\ell} + C_{\ell'})/2$, where, again, $C$ is either $S$ or $N$. \citet{DESY3_2021} argues that this choice is best motivated under the NKA, and also has the advantage of avoiding numerical issues for cross-spectra that can be negative (e.g., as is the case for the geometric mean and the $TE$ spectrum). We combine this with Equation \ref{eq: INKA} when substituting spectra into Equation \ref{eq: general_pseudocov}.

The preceding prescription defines the following required inputs:
\begin{itemize}
    \item Couplings for each spin and combination of four effective masks (for the pseudospectrum covariance matrix).
    \item Couplings for each spin and combination of two effective masks (for the INKA mode-coupling matrices).
    \item $w_2$ factors for each combination of two effective masks (for the INKA).
    \item Fiducial signal and noise power spectra.
\end{itemize}
We briefly elaborate on these elements in the following.

\subsubsection{Coupling Matrices} \label{sec: pipeline_analytic_couplings}

We calculate all unique couplings required for the pseudospectrum covariance, as described above. While the combinatorics for the two-mask couplings are manageable, they become prohibitive for the four-mask couplings if computed naively, with almost 2 million possible combinations. To conserve resources, we tabulate the number of unique couplings (e.g., given that the DR6 analysis masks do not depend on polarization). Then, we discard couplings that would correspond to a cross-split noise power spectrum in Equation \ref{eq: general_pseudocov}: since we assume splits have independent noise, these terms would be zero regardless of the coupling. This results in only 3,062 couplings to compute. Finally, we accelerate computation by using the Toeplitz approximation with the same default parameters as \citet{Toeplitz}. We find errors incurred by this approximation are sub-percent. In parallel, we calculate all required $w_2$ factors at negligible cost.

\subsubsection{Fourier-space Filter Treatment} \label{sec: pipeline_analytic_fourier}

As discussed in \S\ref{sec: formalism_approx_data_model}, we model the $\mathcal{O}(1)$ effect of the Fourier-space filter on the power spectrum covariance. We do this by applying an isotropic transfer function, $t_{\ell}^{\alpha}$, to the fiducial signal and noise power spectra (Equation \ref{eq: approx_data_model}). The filter has a different effect depending on whether the fiducial spectrum appears in the power spectrum estimator or the pseudospectrum covariance matrix, which we model by an isotropic transfer function with different shapes. We capture the different shapes by modifying the exponent of the transfer function, $\alpha$. In other words, when the filtered power spectrum appears in the MASTER pseudospectrum estimator, we multiply it by the ``two-point" transfer function given by $t_{\ell}^{\alpha_{2pt}}$. When the filtered power spectrum appears as part of the NKA (or INKA) in the covariance matrix, we multiply it by the ``four-point" transfer function, $t_{\ell}^{\alpha_{4pt}}$. In this section, we summarize how we determine each component of this model: $t_{\ell}$, $\alpha_{2pt}$, and $\alpha_{4pt}$. A detailed description of each step is given in Appendix \ref{apx: covmat_kspace}.

We first obtain the transfer function template, $t_{\ell}$, by applying the Fourier filter to a small number (50) of full-sky, white-noise simulations. Because the power spectrum for this field is a known constant, any observed deviations in the power spectra of the simulations are due to the filter, independent of any masking. We apply a fourth-order Savitzky-Golay filter to the simulated spectra to construct a smooth template. 

\begin{figure}
    \centering
    \includegraphics[width=\columnwidth]{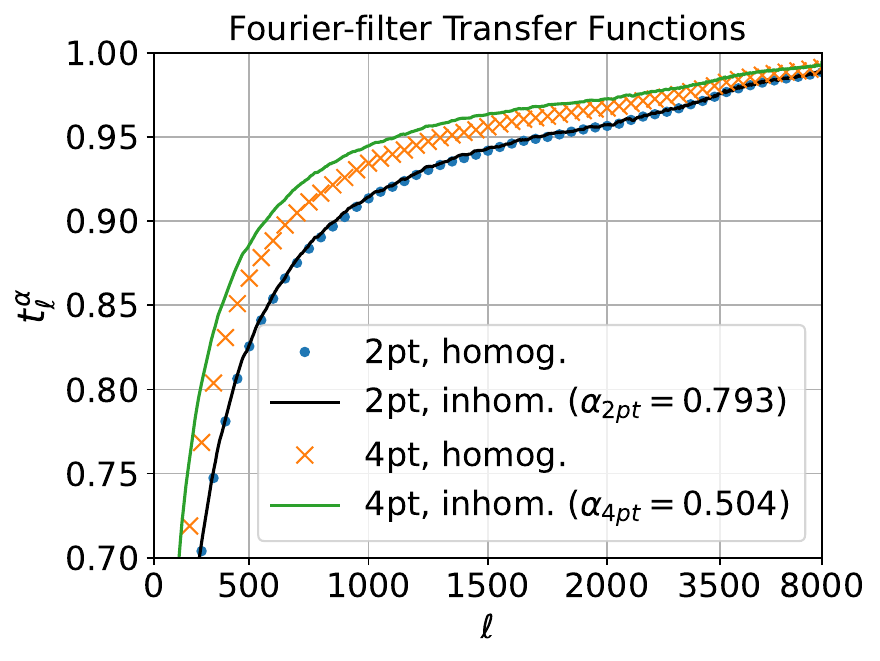}
    \caption{The ``two-point" and ``four-point" Fourier-space filter transfer functions have different shapes. Comparing our new method as part of the inhomogeneous matrix to the homogeneous matrix approach, the two methods agree on the two-point transfer functions to $<1\%$ but disagree on the four-point transfer functions on medium and large scales. The homogeneous matrix transfer function is determined by bin rather than by $\ell$. This figure shows the temperature case.}
    \label{fig: form_kspace_tf}
\end{figure}

Next, we fit for the two-point ($\alpha_{2pt}$) and four-point ($\alpha_{4pt}$) shape exponent. We do so by running a larger number (500) of simulations, which are realizations from a mock, noise-like power spectrum (one for temperature and one for polarization). These simulations are filtered, and their pseudospectra measured using a representative, average effective mask. In the two-point case, we fit for $\alpha_{2pt}$ by forward modeling the observed pseudospectra as $\tilde{C}_{\ell}(\alpha_{2pt}) = \sum_{\ell_1}M_{\ell\ell_1}t_{\ell_1}^{\alpha_{2pt}}C_{\ell_1}$, and analogously for $\alpha_{4pt}$ by forward modeling $\tilde{C}_{\ell}(\alpha_{4pt})$ in an expression for the power spectrum covariance matrix (see Appendix \ref{apx: covmat_kspace}). Our choice to use a mock power spectrum, and single representative effective mask, is justified by our finding that both $\alpha_{2pt}$ and $\alpha_{4pt}$ are largely independent of the mask and power spectrum; instead, they are mainly a function of the actual Fourier-filter. Thus, we can reuse the model for both signal and noise, as well as across arrays and splits. This feature drastically limits the computational cost of these dedicated simulations in comparison to the full-dataset ensemble in \S\ref{sec: pipeline_sims}. Results of this process are shown in Figure \ref{fig: form_kspace_tf} for the polarization case.

The ACT DR4 analysis \citep{C20} also used simulations to determine the approximate two-point transfer function from the Fourier-filter, although the specific implementation differed from this paper. However, as roughly mapped onto our method, \citet{C20} used the fixed relation $\alpha_{4pt}=0.75\alpha_{2pt}$. While a reasonable approximation for our specific Fourier-filter, we find this relation does not generally hold. A comparison of our new model to that of \citet{C20} (which is also used in the homogeneous matrix) is shown in Figure~\ref{fig: form_kspace_tf}, where the difference can be seen on large scales for the four-point transfer function (note, $0.75\alpha_{2pt} = 0.595$, whereas we find $\alpha_{4pt}=0.504$). Because the covariance matrix is quadratic in the power spectra, our new method increases the analytic covariance of each spectrum by $\sim5\%$ at $\ell=500$. In either case, inaccuracies in the Fourier-filter treatment are corrected using the Monte Carlo covariance matrix discussed in \S\ref{sec: pipeline_sims}.

Finally, we note that our procedure for fitting $\alpha_{2pt}$ and $\alpha_{4pt}$ does not \textit{exclusively} capture the effect of the Fourier-space filter. Rather, it can capture \textit{any} deviations from the expected covariance under the MASTER framework. Thus, while targeting the anisotropic Fourier-filter, the model may also capture deviations due to a breakdown of the NKA. It is important to bear this in mind when we evaluate the performance of the ``analytic'' covariance matrix: any purely NKA-induced errors may be suppressed by the Fourier-filter correction. 

\subsubsection{Fiducial Signal Power Spectra} \label{sec: pipeline_analytic_signal}

We use a fiducial signal power spectrum including the CMB, extragalactic foregrounds, and dust, following the cosmology and foreground model of \citet{C20}. Thus, the signal cross-power spectrum of a given pair of arrays is a function of the array passbands (which we take to be fixed) and the spectral energy distributions (SEDs) of each component. The parameters of the model used in this paper are given in Table \ref{tab: signal_model} in Appendix \ref{apx: covmat_signal}. Following \citet{Carron2013, Planck_V_2018}, we assume the fiducial signal power spectra are sufficiently close to the true power spectra that the ACT DR6 cosmological results are unbiased. We then apply the beam transfer functions for each array, and, following \S\ref{sec: pipeline_analytic_fourier}, we account for the four-point effect of the Fourier-space filter by applying $t_{\ell}^{\alpha_{4pt}}$. Finally, following the INKA, we convert the power spectra into pseudospectra using the mode-coupling matrix for the two arrays' effective signal masks.

\subsubsection{Fiducial Noise Power Spectra} \label{sec: pipeline_analytic_noise}

We measure the noise spectra directly from the data. Following \citet{mnms}, we assume uncorrelated noise between maps from different physical detector wafers, but possibly correlated noise between frequency channels on the same wafer. For example, we take PA5 f090 noise to be independent from PA6 f090 noise, but not from PA5 f150 noise. We perform the following steps for each detector wafer, and for each frequency-polarization pair on that wafer. First, we compute the average pseudospectrum over pairs of different map splits (for four splits, there are 12 such pairs).\footnote{The average over 12 pairs is correct regardless of whether the frequency-polarization legs in the pair are different or the same. In the latter case, it is true that we double computation, but our code is simpler and easier to read.} Because the ACT DR6 map splits contain independent noise, this average represents an unbiased signal-only model. We then compute the pseudospectrum for each map split paired with itself (i.e., an ``auto-split" pseudospectrum). This second quantity contains the same signal as the first average in expectation, but with an additive noise bias equal to the noise power spectrum in that split. Subtracting the signal-only model from the ``auto-split" pseudospectrum results in a measurement of the noise-only model. We then reduce statistical scatter by filtering the measured noise pseudospectra with a fourth-order Savitzky-Golay filter. 

Because the data from which we measure the noise pseudospectra have been filtered with the Fourier-space filter, they represent mode-coupled noise power spectra after applying the \textit{two-point} Fourier-filter transfer function. However, following \S\ref{sec: pipeline_analytic_fourier}, the covariance matrix recipe calls for mode-coupled noise power spectra after applying the \textit{four-point} Fourier-filter transfer function. Thus, we first apply the inverse of the mode-coupling matrix for the two arrays' and splits' effective noise masks and then divide-out $t_{\ell}^{\alpha_{2pt}}$ to get ``unfiltered" noise power spectra. Then, like the signal spectra, we apply $t_{\ell}^{\alpha_{4pt}}$ and convert back into noise pseudospectra using the same mode-coupling matrix.

\subsubsection{Pseudospectrum to Power Spectrum Covariance} \label{sec: pipeline_analytic_pseudo2spec}

The preceding ingredients produce pseudospectrum covariance matrix blocks at the array, split, and polarization level via Equation \ref{eq: general_pseudocov}. To convert to the power spectrum covariance matrix, we apply the matrix operation that transforms from pseudospectra to power spectra. For a given measured pair of arrays ($IJ$) and polarizations ($XY$), the power spectrum estimator (in symbolic form) is given by:
\begin{equation} \label{eq: pseudo2spec_2pt}
    \begin{aligned}
        \hat{\mathbf D}_{IJ}^{XY} &= \mathbf U_{IJ}^{XY} \mathbf P \mathbf M_{IJ}^{-1, XY} \frac{1}{N_{\{ij\}}} \sum_{\{ij\}} \hat{\tilde{\mathbf C}}_{I_i J_j}^{XY} \\
        &\equiv \mathbf Q_{IJ}^{XY} \sum_{\{ij\}} \hat{\tilde{\mathbf C}}_{I_i J_j}^{XY}.
    \end{aligned}
\end{equation}
Reading from right to left, $\hat{\tilde{\mathbf C}}_{I_i J_j}^{XY}$ is the vector over $\ell$ of the pseudospectrum for the split-pair $ij$, and the sum is over the $N_{\{ij\}}=12$ split-pairs for which $i\neq j$. As in \S\ref{sec: pipeline_analytic_noise}, this aspect of the power spectrum pipeline ensures the estimator contains no noise bias, regardless of the noise model \citep{noise_bias1, noise_bias2}. The matrix $\mathbf Q_{IJ}^{XY}$ is composed\footnote{\mbf{Q} nearly exactly reproduces the ACT DR6 power spectrum pipeline. It omits, however, some small, additive corrections made to the power spectrum that cannot be easily incorporated into \mbf{Q}, see \S\ref{sec: pipeline_sims}.}  of the inverse mode-coupling matrix, $\mathbf M_{IJ}^{-1, XY}$, a ``weighting-binning" matrix, \mbf{P}, which transforms $C_{\ell} \rightarrow D_{\ell}$, with $D_{\ell} \equiv \ell(\ell+1) / (2\pi) C_{\ell}$ and bins the resulting \mbf{D} vector over $\ell$, and $\mathbf U_{IJ}^{XY}$, which is the power spectrum pipeline's implementation of the two-point Fourier-filter correction. This correction is more precise than that described in \S\ref{sec: pipeline_analytic_fourier}: it is based on more simulations and occurs at the binned power spectrum level, as opposed to the unbinned power spectrum level. We calculate \mbf{Q} explicitly. Because the covariance is bilinear, we have for a block of the binned power spectrum covariance matrix, $\Sigma_{bb'}$:
\begin{align} 
\label{eq: pseudo2spec_4pt}
    \Sigma_{IJ,PQ,bb'}^{WX,YZ} = \sum_{\ell}Q_{b\ell}\sum_{\ell'}Q_{b'\ell'}\sum_{\{ij\}}\sum_{\{pq\}} \tilde{\Sigma}_{I_i J_j,P_p 
    Q_q,\ell\ell'}^{WX,YZ}.
\end{align}
Equation \ref{eq: pseudo2spec_4pt} defines our ``analytic" covariance matrix. 

For convenience, we briefly define a more-compact notation of a covariance matrix block. We combine the polarization pair $WX$ and array pair $IJ$ defining the block of the power spectrum vector into one index, $\beta$. Thus, we may also refer to any element of the binned covariance matrix as $\Sigma_{\beta\beta',bb'}$. We define the \textit{block-wise} matrix diagonal to be the set of matrix blocks $\beta'=\beta$ for any bins, and the \textit{bin-wise} matrix diagonal to be the bins $b'=b$ for any block; the \textit{main} matrix diagonal is the set of elements for which $\beta'=\beta$ and $b'=b$. We may also identify series of off-diagonal blocks (e.g., $\beta'=\beta+1$) and off-diagonal bins (e.g., $b'=b+1$) explicitly.

\subsubsection{Comparison to Previous Methods} \label{sec: pipeline_analytic_comparison}

This analytic covariance is closely related to previous analyses' prescriptions. The analytic part of the ``homogeneous" ACT matrix uses the public power spectrum covariance matrix code \texttt{pspy}, that like \texttt{NaMaster}\footnote{\url{https://github.com/LSSTDESC/NaMaster}} provides convenient functions that assume homogeneous noise weighting (i.e., $v(x)\equiv w(x)$ in Equation \ref{eq: general_data_model}). \citet{Li2023} found that neglecting survey depth in the effective noise masks underestimates \textit{Planck} covariance matrices by $\sim10\%$, while \citet{mnms} found such prescriptions underestimate ACT DR6 covariance matrices by up to $\sim20\%$. In addition to assuming homogeneous survey depth, the homogeneous matrix uses the standard NKA, the ACT DR4-style Fourier-space filter correction (see the discussion in \S\ref{sec: pipeline_analytic_fourier}), a simpler fiducial noise power spectrum pipeline that averages the spectra over split-crosses and uses binning instead of an $\ell$-dependent filter to smooth the measurements, and it sets cross-array noise spectra to be positive and cross-polarization noise spectra to be 0. Further implementation details for the homogeneous matrix are available in the \texttt{pspy} documentation.\footnote{\url{https://pspy.readthedocs.io/en/latest/scientific_doc.pdf}}

The closest analytic covariance prescription to our ``inhomogeneous" matrix is in \citet{Planck_XI_2015}, where the inhomogenous survey depth is accounted for in the effective noise masks, and thus the distinct signal and noise coupling terms are included as in Equation \ref{eq: general_pseudocov}. For treating the signal spectra, \citet{Planck_XI_2015} used the traditional NKA; for the noise, an ``approximate" treatment of its non-white character is noted. Close inspection of \citet{Planck_XV_2013} and \citet{Planck_XI_2015} shows this treatment was equivalent to the INKA, and thus no more approximate than their signal treatment under a MASTER-like data model. Finally, that analysis uses a mix of arithmetic and geometric symmetry of spectra, unlike our use of only arithmetic symmetry, although this difference is minor. The main difference between \citet{Planck_XI_2015} and this paper are in the properties of the ACT data in comparison to \textit{Planck}. The \textit{Planck} noise power spectra, though not perfectly white, are broadly smooth and within $\sim30\%$ of white for scales smaller than $\ell\sim$ 500 (200) for temperature (polarization) \citep{Planck_XI_2015, Li2023}. Due to the atmosphere, the ACT noise power spectra are steeper, and their $1/f$ character persists to smaller scales, especially in temperature. Additionally, \textit{Planck} noise adheres more closely to the MASTER data model, lacking the spatially-dependent stripy noise patterns that are prominent in ACT \citep[see e.g.,][]{N20}. 

\subsection{Simulations} \label{sec: pipeline_sims}

We complement the analytic power spectrum covariance matrix of \S\ref{sec: pipeline_analytic} with a Monte Carlo covariance made from a full-dataset simulation ensemble. We generate this ensemble of simulations following the full ACT data model of Equation \ref{eq: full_data_model} and Appendix \ref{apx: sims}. This Monte Carlo covariance serves as a noisy estimate of an unbiased covariance matrix that incorporates the full data realism.

The signal components of the simulations are drawn from the same fixed signal model as in \S\ref{sec: pipeline_analytic}. The processing of the signal components includes an additive correction for $TE$ power spectra due to the Fourier-filter that cannot be incorporated into the \mbf{U} matrices from \S\ref{sec: pipeline_analytic_pseudo2spec}. As a baseline, we draw the noise components from the tiled noise model (see Equation \ref{eq: noise_def}). Between the directional wavelet model and the tiled model, only the tiled model supports simulations beyond the smallest scale of the DR6 power spectrum ($\ell_{max}=8,500$). In particular, the noise simulations are drawn at half the resolution of the data --- 1 arcmin pixels --- supporting scales as small as $\ell_{max}=10,800$. Further details are provided in Appendix \ref{apx: mnms} and \citet{mnms}.

Including signal and noise, in our baseline ensemble we draw 1,600 realizations of each array, polarization, and split. Each simulation is passed through the power spectrum reconstruction pipeline, and we use the ensemble of simulated spectra to construct the Monte Carlo covariance. We note that the simulations do not include two small corrections made to the data: beam leakage, because its uncertainty is propagated into the covariance analytically; and aberration due to Earth's motion with respect to the CMB rest frame, because this is a second-order effect at the covariance level. This ensemble is sufficient to constrain each element of the covariance matrix to $\lesssim 3.5\%$; nevertheless, drawing enough simulations to ensure a well-conditioned, non-singular Monte Carlo matrix is not feasible.

\subsection{Simulation-Based Correction} \label{sec: pipeline_sim_correction}

The goal of the covariance pipeline is to produce an accurate, yet tractable, covariance matrix for the ACT DR6 power spectrum. While the Monte Carlo matrix is not limited by the approximations of the analytic matrix, drawing enough simulations to reach convergence is prohibitively expensive. Thus, we start by assuming the analytic matrix is a good, if slightly biased, estimate of the true covariance matrix. Then, we use the Monte Carlo matrix to correct the analytic matrix where they measurably differ. This procedure --- optimizing between a biased, analytic covariance estimate and a noisy, unbiased Monte Carlo covariance estimate --- is referred to as matrix ``shrinkage" or ``conditioning" and is a well-studied problem in applied statistics \citep[see e.g.,][]{Ledoit2003, Schafer2005, Ledoit2020}. Recently, \citet{Balkenhol2022} and \citet{Looijmans2024} reviewed several covariance matrix shrinkage methods for cosmology. In this paper, we opt for a new, simple shrinkage method.

Our method relies on two core assumptions: the eigenbases of the analytic and Monte Carlo covariances are close, and the ratio of their eigenspectra are smooth. Defining the matrix exponent of the covariance as:
\begin{align}
    \bm\Sigma^{p} \equiv \mathbf O \mathbf E^p \mathbf O^T,
\end{align}
where \mbf{O} is orthogonal and \mbf{E} is diagonal, we can construct a \textit{rotated} Monte Carlo covariance matrix:
\begin{equation} \label{eq: Sigma_R}
    \begin{aligned}
        &\bm\Sigma_R \equiv \bm\Sigma_A^{-\frac{1}{2}}\bm\Sigma_M\bm\Sigma_A^{-\frac{T}{2}} \\
        &= \mathbf O_A \mathbf E_A^{-\frac{1}{2}} (\mathbf O_A^T \mathbf O_M) \mathbf E_M (\mathbf O_{M }^T \mathbf O_A) \mathbf E_A^{-\frac{1}{2}} \mathbf O_A^T \\
        &\approx \mathbf O_A (\mathbf E_M \mathbf E_A^{-1}) \mathbf O_A^T,
    \end{aligned}
\end{equation}
where $M$~($A$) denotes the Monte Carlo (analytic) matrix, and we have used $\mathbf O_A \approx \mathbf O_M$. Then, $\mathbf E_M \mathbf E_A^{-1}$ is a diagonal matrix populated by the ratio of the Monte Carlo and analytic eigenspectra. If this ratio is approximately flat, then $\bm\Sigma_R$ is close to diagonal, and its elements preserve the original ordering of the data vector. 

\begin{figure}
    \centering
    \includegraphics[width=\columnwidth]{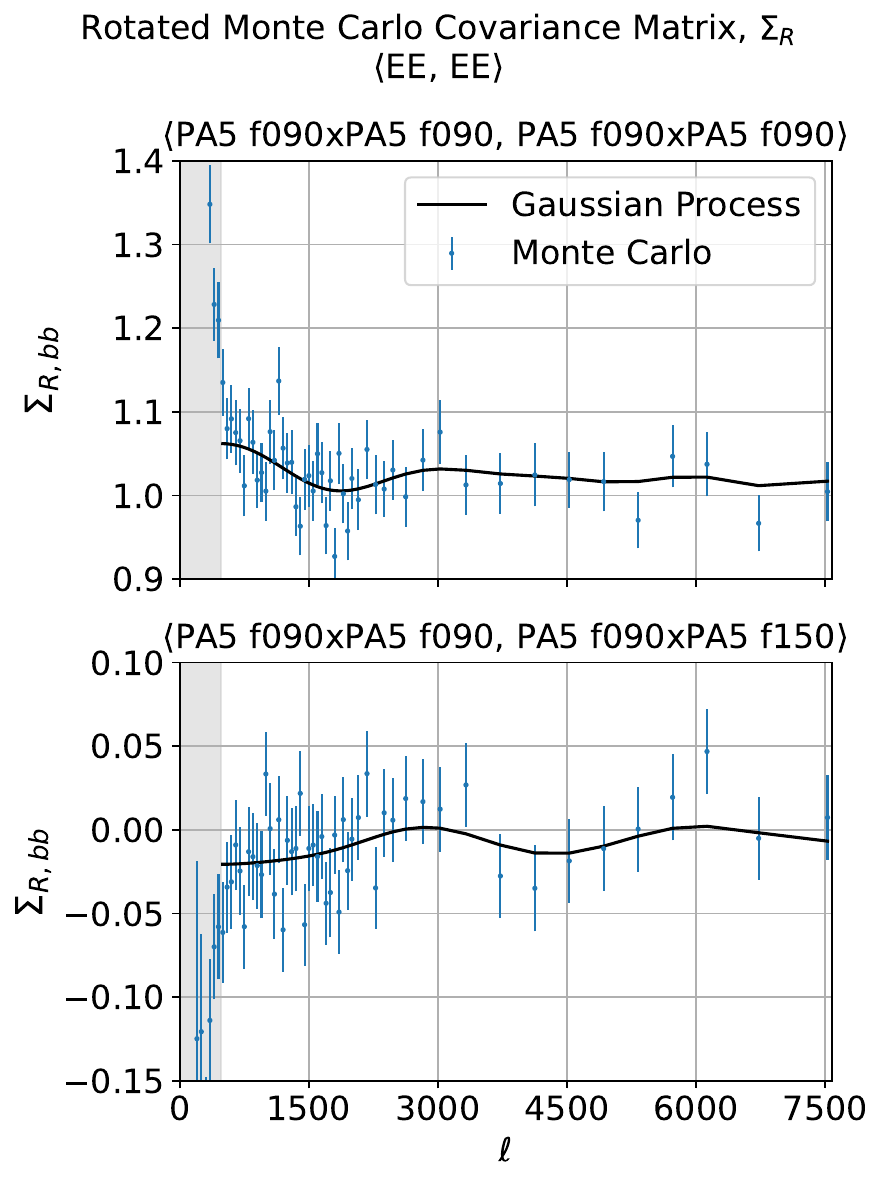}
    \caption{\textit{Top:} The main diagonal of the $\bm\Sigma_R$ matrix defined in Equation \ref{eq: Sigma_R}, for the PA5 f090 x PA5 f090 $EE$ block. If the analytic and Monte Carlo covariance matrices are sufficiently close, this is approximately the ratio of the eigenvalues of the two matrices. \textit{Bottom:} The bin-wise diagonal of the covariance between the PA5 f090 x PA5 f090 $EE$ and PA5 f090 x PA5 f150 $EE$ blocks. In both cases, the $1\sigma$ scatter of the Monte Carlo estimates are shown, with a Gaussian process fit using data above the scale-cut, indicated by the grey region. The Gaussian processes do not account for correlations between bins.}
    \label{fig: form_GP}
\end{figure}

We find these approximations perform well, but are not exact. Specifically, while $\bm\Sigma_R$ is diagonally-dominant with diagonal values of $\mathcal{O}(1)$, the bin-wise diagonals ($b'=b$) of its off-diagonal blocks ($\beta'\neq\beta)$ are non-zero at the few-percent level. This motivates the following correction procedure: for each block (i.e., for each $\beta\beta'$ pair), we extract the corresponding bin-wise diagonal of $\bm\Sigma_R$ and smooth it using a Gaussian process with a radial basis function (RBF) kernel. The fit involves optimizing two hyper-parameters of the Gaussian process --- the characteristic amplitude and length scale of any coherent features in the data --- given the observed data and its errors. In this case, ``data" refers to the values of the $\bm\Sigma_{R}$ diagonal, and Gaussian errors on $\bm\Sigma_{R}$ are estimated directly from the scatter of the simulations. Example Gaussian process fits are shown in Figure \ref{fig: form_GP}. We only perform the fit over $\ell$-bins preserved by our pre-unblinding scale cuts, discussed in \S\ref{sec: data}.\footnote{The noise simulations entering the Monte Carlo covariance matrix have excess power at scales larger than the scale cuts, leading to the discrepant large values at low $\ell$ in Figure \ref{fig: form_GP}, that would otherwise bias the fits. Also see the discussion in \S\ref{sec: res_mc_validation}.} Finally, for each block, we set all bin-wise off-diagonal elements ($b'\neq b$) to zero. This procedure yields the ``corrected" matrix, $\bm\Sigma_{R,corr}$. Finally, we rotate back to the original basis, yielding the final corrected covariance matrix:
\begin{align}
    \bm\Sigma_{corr} \equiv \bm\Sigma_A^{\frac{1}{2}}\bm\Sigma_{R,corr}\bm\Sigma_A^{\frac{T}{2}}.
\end{align}
This procedure has no free parameters and is fast to compute.

\section{Results} \label{sec: results}

\begin{figure*}
    \centering
    \includegraphics[width=0.85\textwidth]{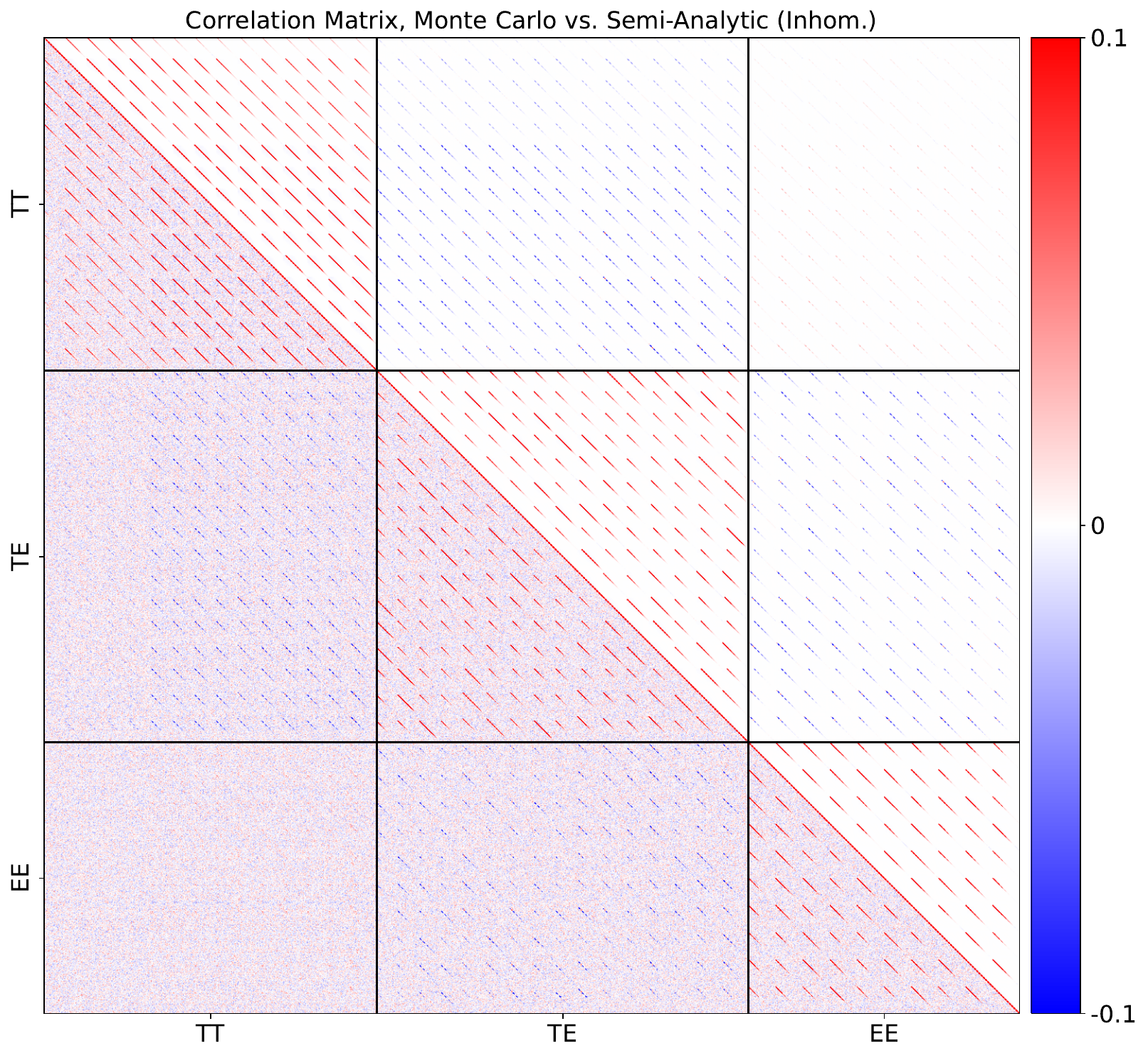}
    \caption{\textit{Below-left of main diagonal:} The correlation matrix for the Monte Carlo covariance. Most elements are zero-mean statistical noise. \textit{Above-right of main diagonal:} The correlation matrix for the corrected analytic covariance from the inhomogeneous prescription. We apply the data and scale cuts from \S\ref{sec: data}, and restrict to the $TT$, $TE$, and $EE$ polarization pairs that enter the DR6 likelihood. As noted in the text, additional correlation from non-Gaussian sky components and beam uncertainty are discussed in the ACT DR6 power spectrum paper and are not plotted here.}
    \label{fig: res_mc_vs_ana_corrected_mat}
\end{figure*}

Our semi-analytic covariance matrix relies on several approximations unique to the ACT DR6 data. In this section, we evaluate the accuracy of the pipeline by comparison to the Monte Carlo covariance matrix. Since we take the Monte Carlo matrix to be an unbiased estimate of the true covariance, we also probe its internal robustness.

\subsection{Covariance Matrix Pipeline Validation} \label{sec: res_ana_validation}

Here we show that the semi-analytic covariance matrix achieves excellent agreement with the Monte Carlo matrix, and that the size of the corrections applied in \S\ref{sec: pipeline_sim_correction} are small.

\begin{figure*}
    \centering
    \includegraphics[width=0.85\textwidth]{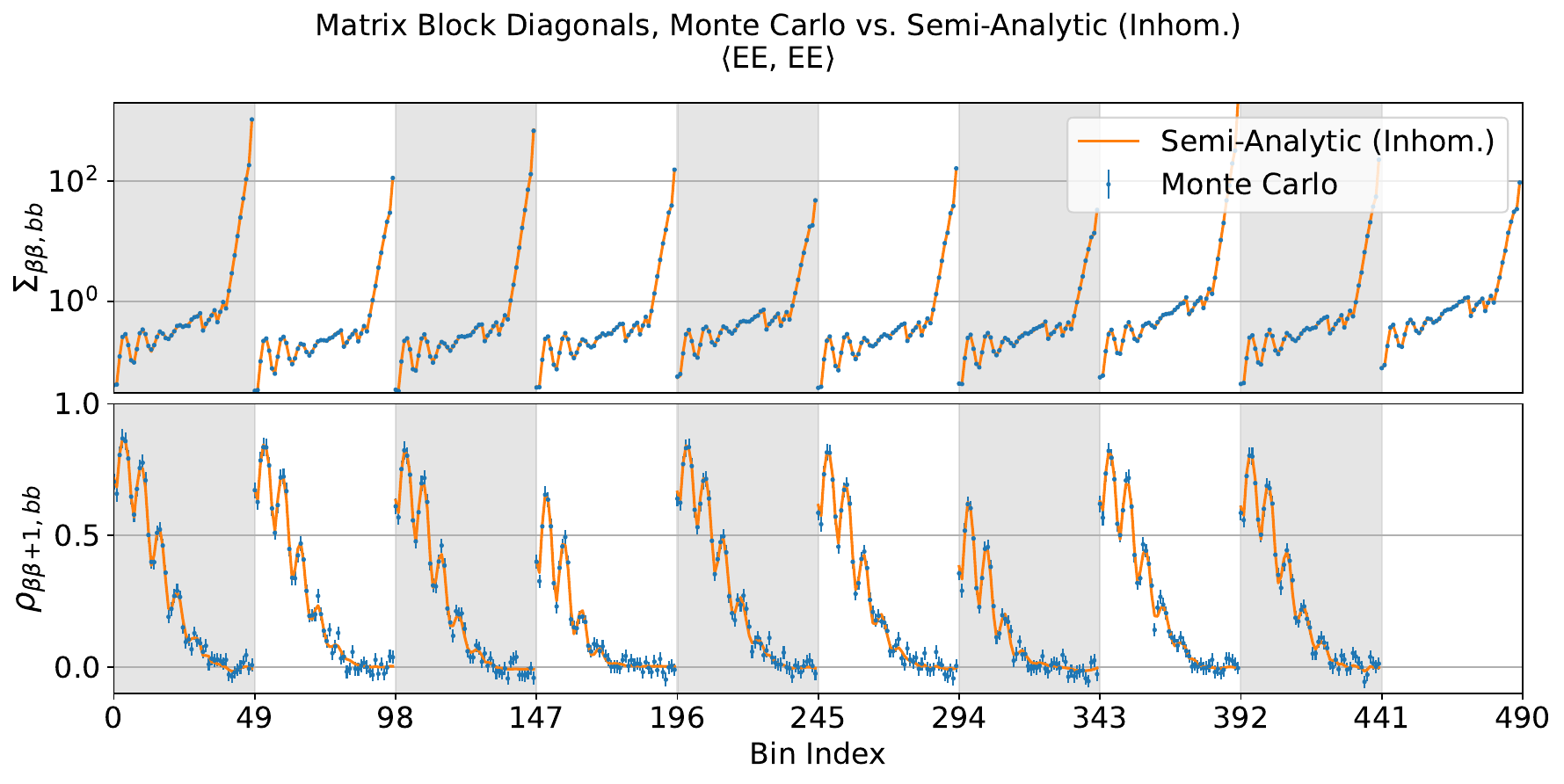}
    \includegraphics[width=0.85\textwidth]{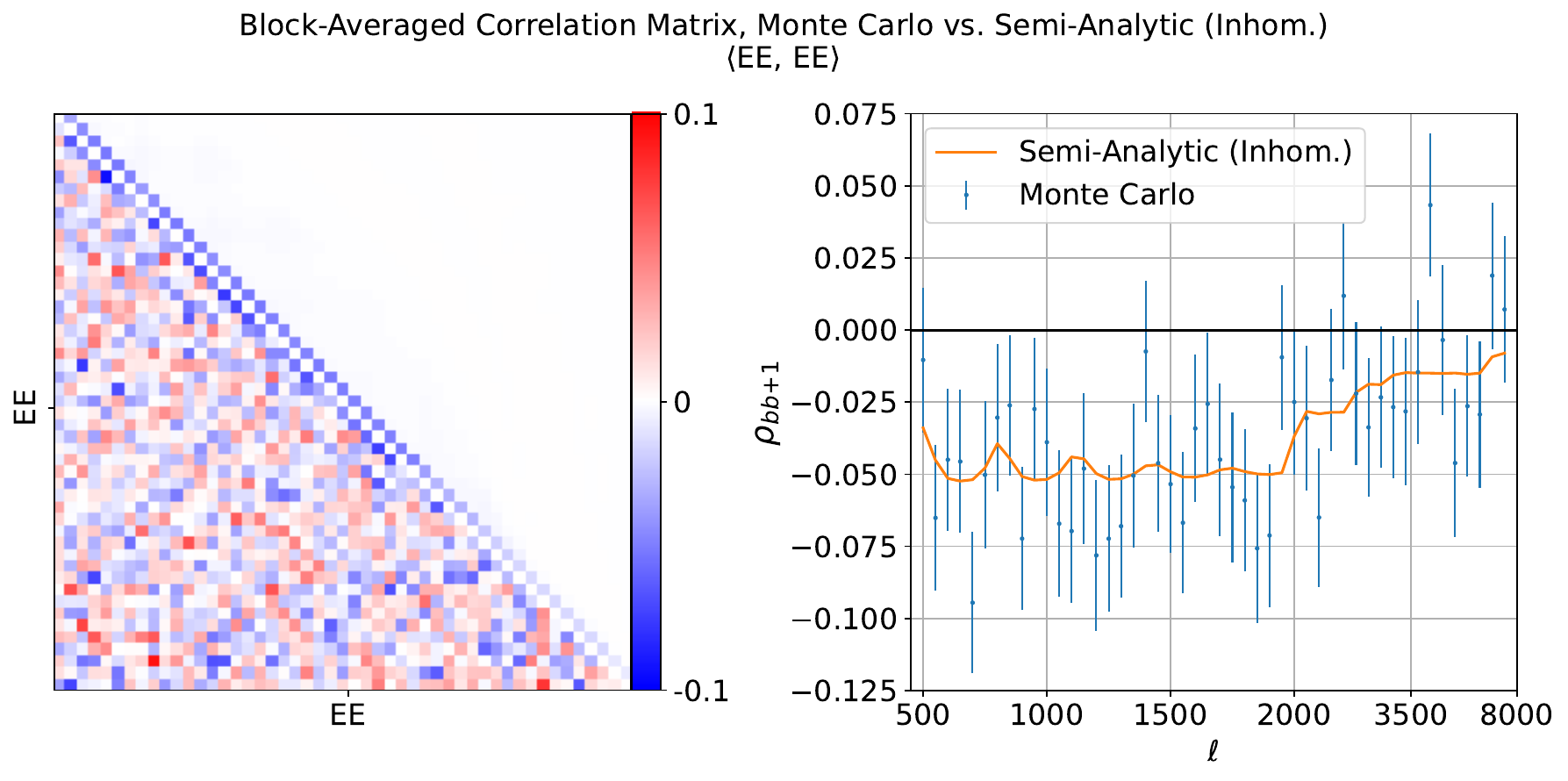}
    \caption{\textit{First row:} Main diagonal ($\beta'=\beta$, $b'=b$) of the $EE$ part of the covariance matrix, $\bm\Sigma$. The data and scale cuts from \S\ref{sec: data} result in 10 $EE$ array-pairs, or blocks, each containing 49 bins. The boundaries of these blocks along the main diagonal are denoted by the alternating grey and white bands. \textit{Second row:} The bin-wise diagonal ($b'=b$) of the first off-diagonal blocks ($\beta'=\beta+1$) of the $EE$ part of the correlation matrix, $\bm\rho$. There are only 9 such blocks. In both cases, the matrix contains significant structure and dynamic range. \textit{Bottom left:} The correlation matrix after averaging over all 100 blocks of the $EE$ part of the covariance matrix. As in Figure \ref{fig: res_mc_vs_ana_corrected_mat}, below-left of the main diagonal shows the Monte Carlo values, and above-right shows the semi-analytic, inhomogeneous matrix result. The diagonal of the correlation matrix has been set to zero to enhance visibility of the off-diagonal elements. \textit{Bottom right:} A direct comparison of the first off-diagonal bins ($b'=b+1$) in the average correlation matrix reveals $\sim5\%$-level correlations in the simulations that are captured by the semi-analytic covariance.} 
    \label{fig: res_mc_vs_ana_corrected_mat_coadd_EE}
\end{figure*}

The structures of both the Monte Carlo matrix and the semi-analytic covariance matrix are dominated by the bin-wise diagonals ($b'=b$) across all diagonal and off-diagonal blocks (all $\beta\beta'$ pairs). We demonstrate this visually by plotting the correlation matrix of both matrices in Figure \ref{fig: res_mc_vs_ana_corrected_mat}. For a generic covariance matrix $\Sigma_{ij}$ with elements indexed by $ij$, the correlation matrix is defined as:
\begin{align} \label{eq: correlation_matrix_def}
    \rho_{ij} \equiv \Sigma_{ij} / \sqrt{\Sigma_{ii}\Sigma_{jj}}.
\end{align}
Figure \ref{fig: res_mc_vs_ana_corrected_mat} shows the similarity between the off-diagonal structures of the Monte Carlo and semi-analytic correlation matrices. We can easily discern the blocks associated with the 15 $TT$, 16 $TE$, and 10 $EE$ cross-array spectra in the data vector by their prominent bin-wise diagonals for each block.

\begin{figure*}
    \centering
    \includegraphics[width=0.85\textwidth]{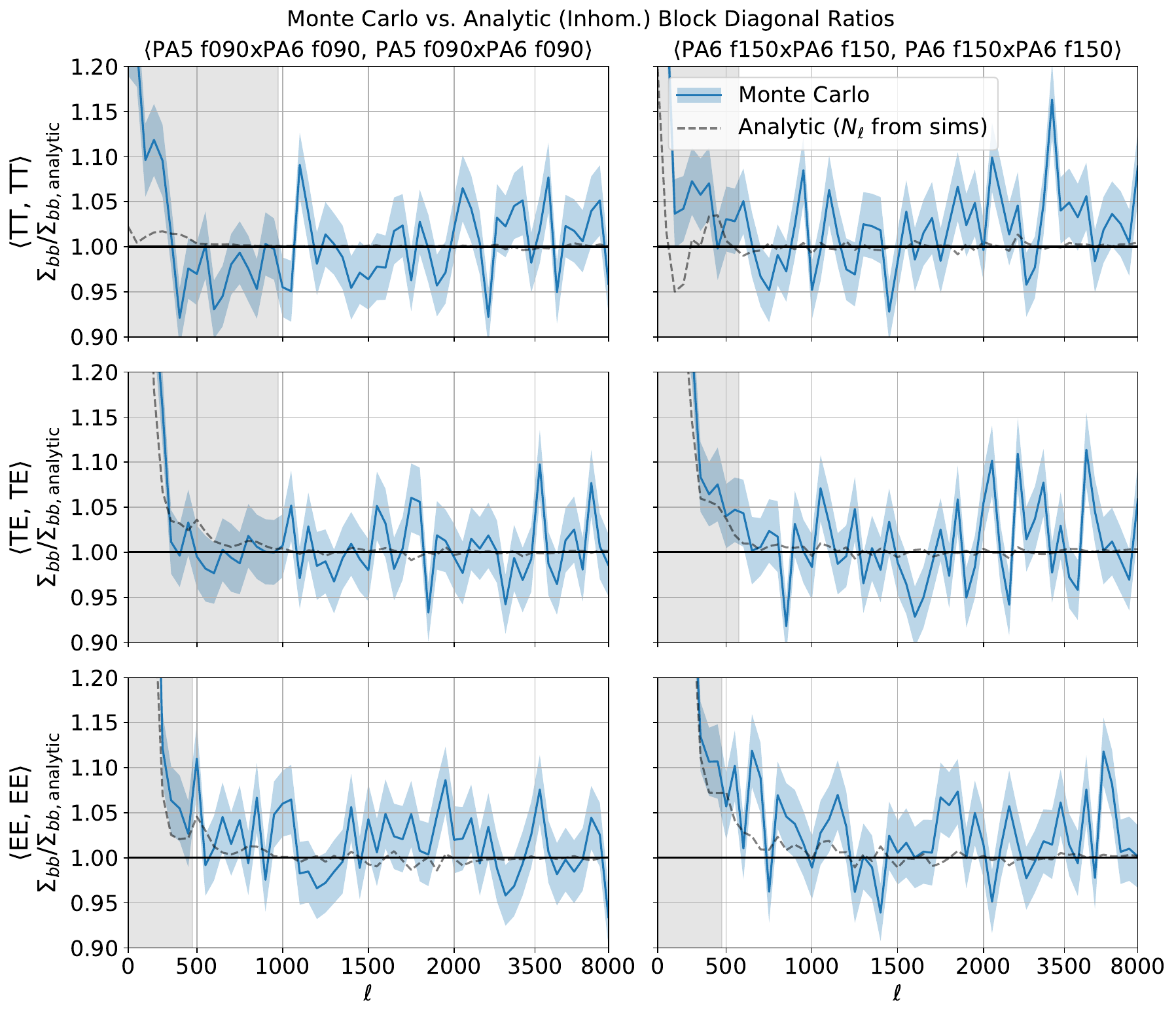}
    \caption{Ratios of the diagonal of the Monte Carlo covariance to the diagonal of the inhomogeneous analytic covariance, for a selection of array and polarization spectra. The PA5 f090 x PA6 f090 $TT$ spectra are the most signal-dominated; the PA6 f150 x PA6 f150 $EE$ spectra are the most noise-dominated. The grey shaded regions represent data not included in the likelihood from the scale cuts. The grey dashed line represents the amount of the observed excess Monte Carlo variance due to excess noise power in the simulations rather than a deficiency of the analytic covariance pipeline. It is an inhomogenous analytic covariance matrix, but substitutes the data with a simulation when measuring the fiducial noise power spectra (see the discussion in \S\ref{sec: res_mc_validation}).}
    \label{fig: res_mc_vs_ana_block_diags}
\end{figure*}

We provide a more detailed view of the matrix structure in Figure \ref{fig: res_mc_vs_ana_corrected_mat_coadd_EE}. The main diagonal of the $EE$ blocks of $\bm\Sigma$ shows considerable dynamic range in simulations, spanning three orders of magnitude. Likewise, the bin-wise diagonals ($b'=b$) of the first off-diagonal $EE$ blocks ($\beta'=\beta+1$) of the correlation matrix also vary from $\sim80\%$ to near-zero correlation depending on the bin. In either case, the structures are well-modeled by the semi-analytic covariance. Finally, we average over all 100 $EE$ blocks in the covariance, elucidating that adjacent bins ($b'=b+1$) are also correlated, though only at the few-percent level. Otherwise, farther bin-wise off-diagonals ($|b'-b|\geq2$) are broadly consistent with noise for all blocks. We quantify that assessment as follows. Assuming zero population correlation, the variance of the Monte Carlo estimator for $\rho_{ij}$ is:
\begin{align} \label{eq: var_correlation_matrix_def}
    \langle\rho_{ij}^2\rangle \approx 1/n_{sim}
\end{align}
in the limit of a large number of simulations, $n_{sim}$ \citep{Hotelling1953}.\footnote{While the Monte Carlo covariance is Wishart distributed, the Monte Carlo correlation is more complicated. However, in this limit, they give the same result.} For $n_{sim}=1,600$, we would expect $\bar{\rho}_{ij}=0$ and $\sigma(\rho_{ij})=2.5\%$. We find $\bar{\rho}_{ij}=-0.01\%$ and $\sigma(\rho_{ij})=2.50\%$ (rounded to two decimals) for the $|b'-b|\geq2$ off-diagonals of the entire correlation matrix (Figure \ref{fig: res_mc_vs_ana_corrected_mat}), consistent with statistical scatter about zero correlation. Accordingly, the semi-analytic matrix predicts negligible correlation for these elements. In summary, the structures of the Monte Carlo and semi-analytic covariance matrices are consistent.

\begin{figure*}
    \centering
    \includegraphics[width=0.85\textwidth]{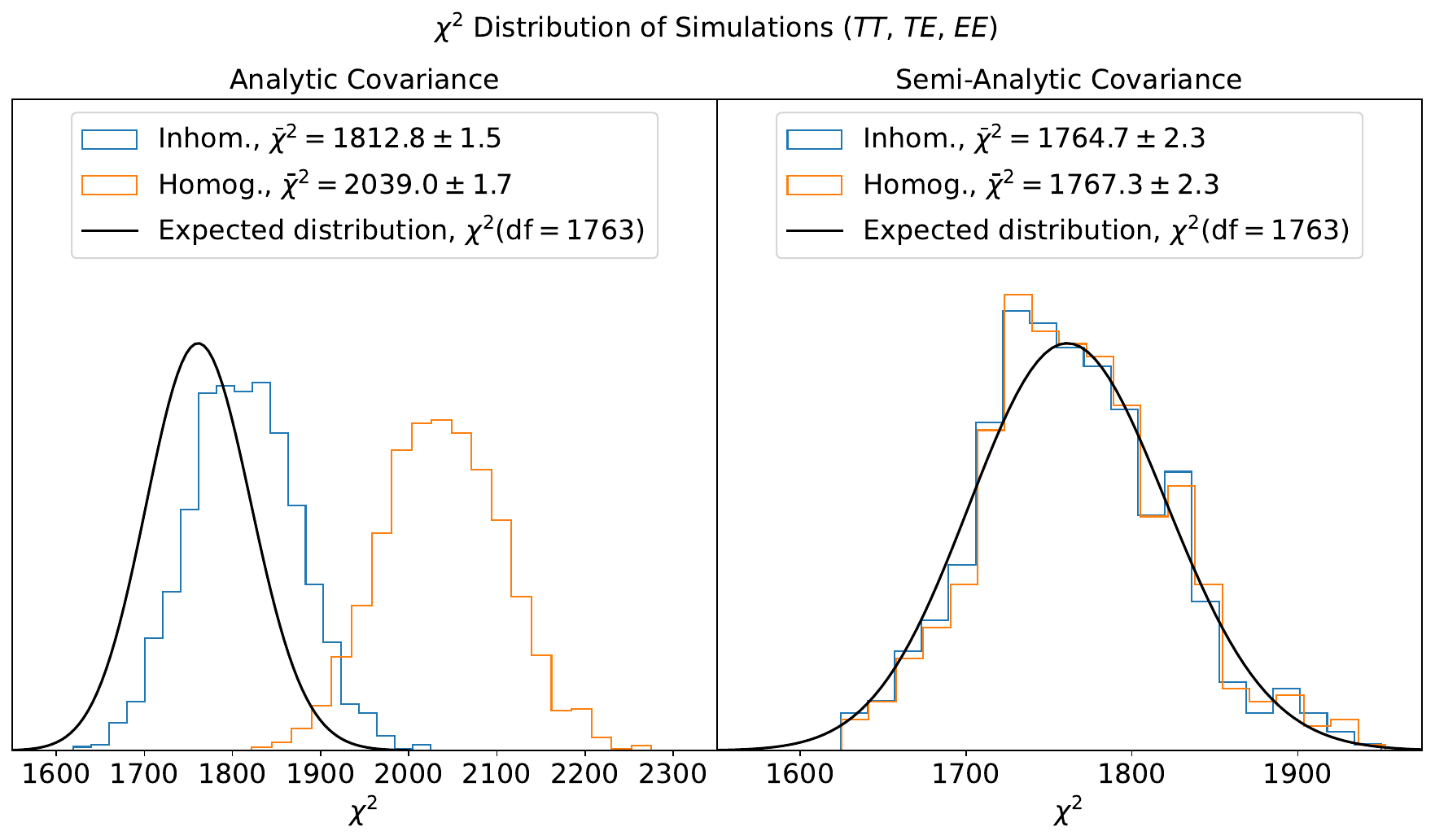}
    \caption{Distributions of the $d_{sim}^2$ (Equation \ref{eq: mahalanobis_def}) statistic for the simulated data vectors using different covariance matrices, after applying scale cuts and restricting to $TT$, $TE$, and $EE$ spectra. \textit{Left:} Using the inhomogeneous and the homogeneous uncorrected analytic covariance matrices, and evaluating all 1,600 simulations in our ensemble. \textit{Right:} After the simulation-based correction (based on 1,000 simulations), and evaluating $d_{sim}^2$ using the remaining 600 simulations. In both cases, the black line gives the expected distribution given the number of degrees-of-freedom (the length of the data vectors).}
    \label{fig: res_chi2_red_ana_old}
\end{figure*}

In addition to their off-diagonals, we compare a subset of the Monte Carlo and analytic matrix diagonals in Figure \ref{fig: res_mc_vs_ana_block_diags}. Here we are evaluating the uncorrected, inhomogeneous analytic covariance that results from \S\ref{sec: pipeline_analytic} alone. We do so to check the magnitude of any pipeline approximations breaking down, before the simulations correct for those approximations. As in Figure \ref{fig: res_mc_vs_ana_corrected_mat_coadd_EE}, we again find percent-level agreement with the Monte Carlo matrix within the DR6 scale cuts. This finding holds across high and low signal-to-noise array-crosses (PA5 f090 x PA6 f090 vs. PA6 f150 x PA6 f150) and polarizations. In comparison to \textit{Planck} \citep[see Appendix C.1.4 of][]{Planck_XI_2015}, we do not find clear evidence for coherent $\sim10\%$-level NKA violations; however, our bin resolution makes this determination difficult. The most significant deviations are seen at large scales in low signal-to-noise polarization blocks. As noted in \S\ref{sec: pipeline_sim_correction}, however, these are due to excess power in the \textit{simulations} entering the Monte Carlo matrix, not a breakdown of the analytic matrix pipeline. We discuss this aspect of the noise simulations further in \S\ref{sec: res_mc_validation} and Appendix \ref{apx: mnms_Nl_excess}. Otherwise, we conclude that the Monte Carlo and analytic matrix diagonals are consistent at the percent-level even prior to the simulation-based correction. For example, of the diagonals shown in Figure \ref{fig: res_mc_vs_ana_block_diags}, the largest mean difference between the Monte Carlo and analytic covariance is $2.6\%$ (PA6 f150 $EE$), of which $0.6\%$ is due to excess power in the simulations.

Lastly, we test the uncorrected and corrected analytic covariance matrix by aggregating the $\chi^2$ distribution of realistic simulations. For each simulation in the ensemble, we compute the squared Mahalanobis distance \citep{Mahalanobis2018} of each simulation:
\begin{align} \label{eq: mahalanobis_def}
    d_{sim}^2 \equiv (\mathbf{\hat{C}}_{sim} - \mathbf C_{th})^T \bm \Sigma^{-1} (\mathbf{\hat{C}}_{sim} - \mathbf C_{th}),
\end{align}
where $\mathbf{\hat{C}}_{sim}$ are the reconstructed data vectors of the simulation, $\mathbf C_{th}$ are the theoretical data vectors assuming the fiducial signal spectra of \S\ref{sec: pipeline_analytic_signal}, and $\bm\Sigma$ is the trial covariance matrix. If the $\mathbf{\hat{C}}_{sim}$ are normally distributed with mean $\mathbf C_{th}$ and covariance $\bm\Sigma$, then $d_{sim}^2$ are $\chi^2$ distributed with degrees-of-freedom given by the size of the vector. This test is comprehensive as $d_{sim}^2$ is sensitive to the entire covariance, and $\mathbf{\hat{C}}_{sim}$ is free from analytic approximations. Results are shown in Figure \ref{fig: res_chi2_red_ana_old}. For the uncorrected analytic covariance, we find our new inhomogeneous matrix leads to a simulation $\chi^2$ distribution that is $\sim2.8\%$ greater than nominal. In other words, on average the uncorrected analytic covariance underestimates the error-bar of simulations by only $\sim1.4\%$. This is a significant finding: despite the ways in which the DR6 data and processing break the assumptions of the MASTER covariance framework discussed in \S\ref{sec: formalism}, an approximate, MASTER-compatible prescription can limit bias to the percent-level. Consistent with the findings of \citet{mnms}, the homogeneous matrix is $\sim15.7\%$ discrepant with simulations. Importantly, after applying the simulation-based correction of \S\ref{sec: pipeline_sim_correction} to each matrix, the resulting $\chi^2$ distributions become consistent with the expected distribution at the $\sim0.3\%$-level.\footnote{To obtain unbiased distributions, the covariance matrix and simulations in Equation \ref{eq: mahalanobis_def} must be statistically independent. We achieve this for the semi-analytic covariance matrices by using only 1,000 simulations to correct the covariance, and the remaining 600 to evaluate the $d_{sim}^2$ statistics. We confirmed that the simulation-based correction is converged for the reduced number of simulations.} Thus, our simulation-based correction method is effective, even for $\sim16\%$-level mismodeling of the input analytic covariance. We conclude that the semi-analytic covariance matrix for ACT DR6, whether using the inhomogeneous or homogeneous prescription, achieves satisfactory performance in the context of the challenging DR6 data properties.

\subsection{Monte Carlo Matrix Validation} \label{sec: res_mc_validation}

The results of \S\ref{sec: res_ana_validation} assume the Monte Carlo covariance represents an unbiased estimate of the true covariance. Here, we investigate the validity of this assumption, focusing on the accuracy of the noise contribution. We briefly compared the recovered noise power spectra between simulations and the data in Figures \ref{fig: form_GP} and \ref{fig: res_mc_vs_ana_block_diags}, noting that the large-scale, polarized noise power spectra in simulations tend to be larger than those of the data. The effect on the covariance is captured by the grey dashed line in Figure \ref{fig: res_mc_vs_ana_block_diags}, showing the  uncorrected analytic covariance where, rather than measuring the fiducial noise power spectrum from the data, we replace the data with a noise simulation in \S\ref{sec: pipeline_analytic_noise}. This accounts for the mismatch between the Monte Carlo and analytic matrix that is due only to the simulation noise power spectra not matching that of the data, rather than due to a breakdown of the covariance pipeline approximations. We see that most of the low-$\ell$ Monte Carlo excess is due to this issue in the noise simulations, and is larger for covariance blocks that are more noise-dominated. This bias has a negligible effect given the DR6 scale cuts; at most, the error-bars of the $\sim$3 largest-scale $EE$ bins above the scale cuts are $\sim1-2\%$ inflated. We discuss the origin of this issue in Appendix \ref{apx: mnms_Nl_excess}, and note it can be improved at larger scales for future analyses.

To attempt to quantify the importance of the spatially-varying anisotropy pattern, which cannot be easily accommodated in the MASTER covariance framework, we compare Monte Carlo matrices constructed from both ACT noise models. This test is well-motivated: the tiled model assumes the noise is diagonal in 2D Fourier space (such that the noise is close to diagonal in spherical harmonic space), whereas the directional wavelet noise model assumes long-range correlations in 2D Fourier space \citep[for more detail, see][]{mnms}. We draw an ensemble of 600 directional wavelet-based simulations, and evaluate their $\chi^2$ distribution against the tile-based semi-analytic covariance matrix, finding good agreement: the wavelet-based ensemble is only $\sim0.7\%$ different for the $TT$, $TE$, and $EE$ spectra. A stronger test selects only for the $TB$, $EB$, and $BB$ spectra, since these spectra are noise-dominated. Notably, we find a similar level of agreement: $\sim0.6\%$.\footnote{In both these cases, we apply the lower bandlimit ($\ell_{max}=5,400$) of the directional wavelet model to both the simulations and the covariance matrix. Doing so to a set of tiled simulations yields the same $\sim0.7\%$ (for $TT$, $TE$, and $EE$) and $\sim0.6\%$ (for $TB$, $EB$, and $BB$) difference, indicating that even these small discrepancies are due to the bandlimit, not the noise model.}  Thus, we cannot detect any significant change in the Monte Carlo covariance, even when allowing for long-range noise correlations over angular scales.

\subsection{Discussion of Semi-Analytic Matrix Performance} \label{sec: discussion} 

\begin{figure*}
    \centering
    \includegraphics[width=0.85\textwidth]{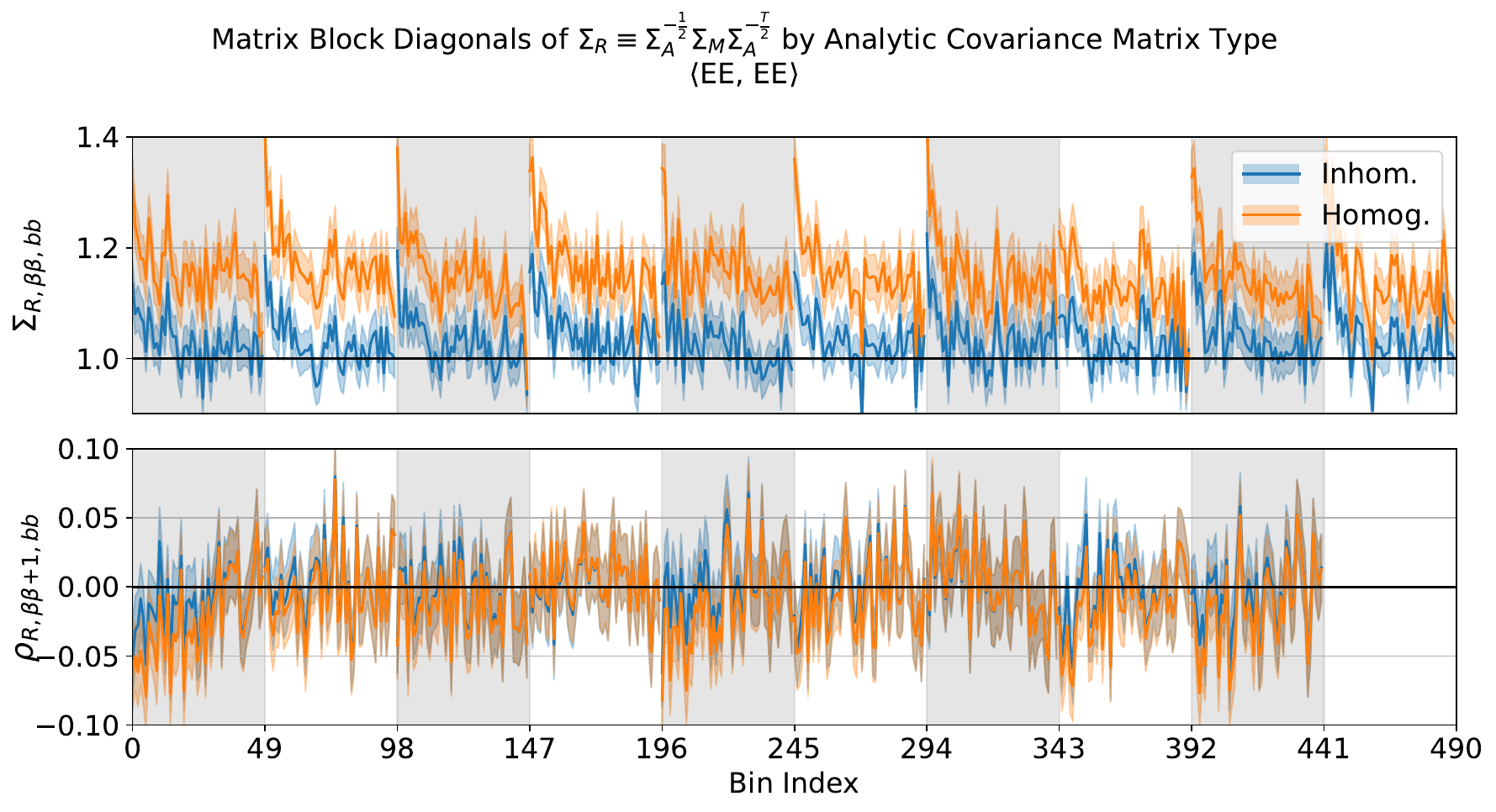}
    \includegraphics[width=0.85\textwidth]{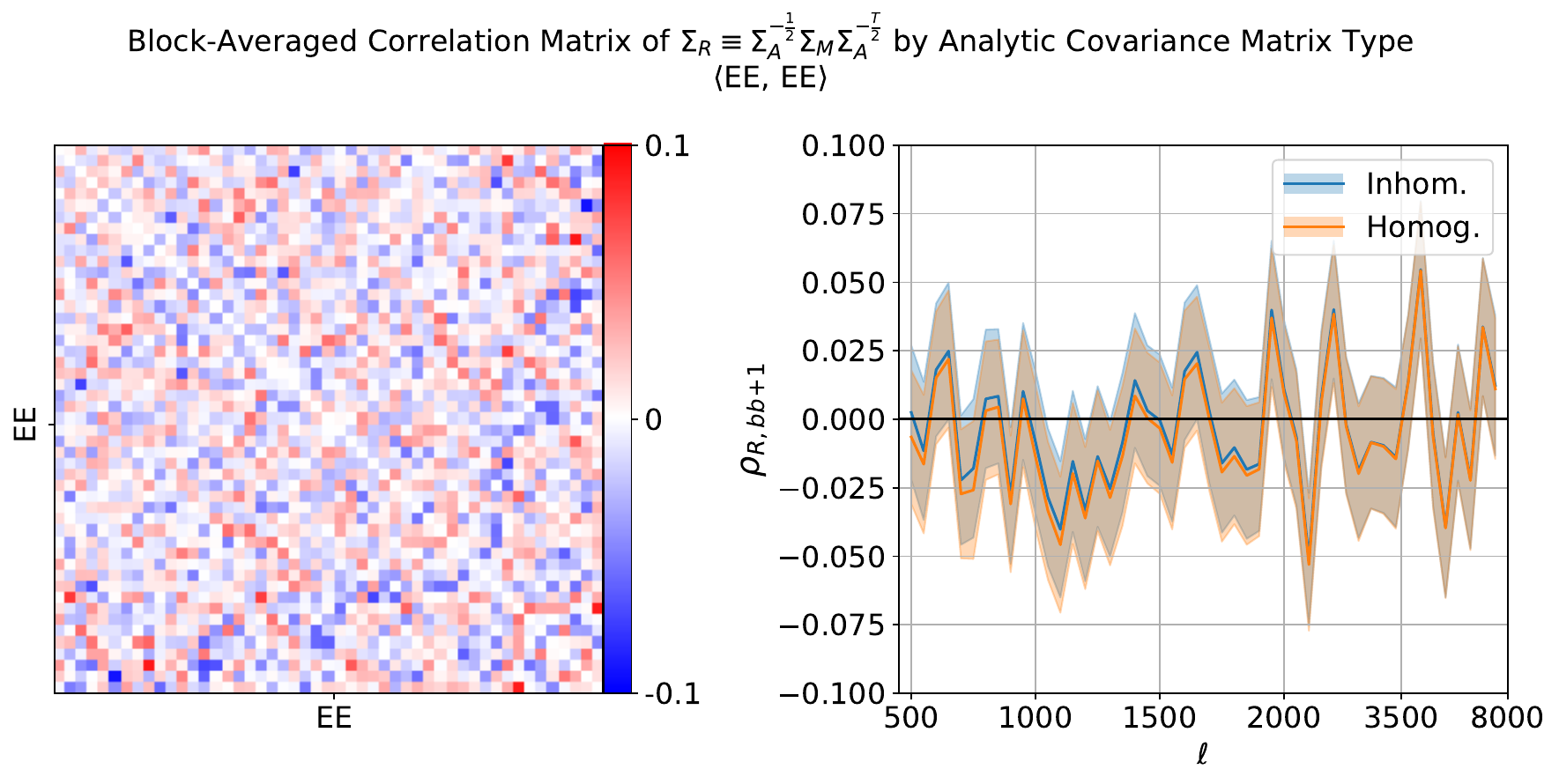}
    \caption{\textit{Top:} Comparison of the rotated Monte Carlo matrices, $\bm\Sigma_R$ (Equation \ref{eq: Sigma_R}), when using the inhomogeneous (blue) or homogeneous (orange) analytic covariance matrices to perform the rotation. The covariance main diagonal elements are close to one, while the correlation bin-wise diagonals are small, but nonzero, for off-diagonal blocks. The matrix values can be directly compared to the first two rows of Figure \ref{fig: res_mc_vs_ana_corrected_mat_coadd_EE}, showing the suppressed structure and dynamic range of $\bm\Sigma_R$ relative to to $\bm\Sigma$. \textit{Bottom left:} The correlation matrix after averaging over all 100 blocks of the $EE$ part of $\bm\Sigma_R$. Below-left (above-right) of the main diagonal shows the result using the homogeneous (inhomogeneous) analytic matrix. The diagonal of the correlation matrix has been set to zero. \textit{Bottom right:} The first off-diagonal bins ($b'=b+1$) in the average $\bm\Sigma_R$ $EE$ correlation matrix are consistent with each other and with zero for each analytic prescription, also unlike in Figure \ref{fig: res_mc_vs_ana_corrected_mat_coadd_EE}.} 
    \label{fig: res_why_does_PSpipe_do_well_coadd_EE}
\end{figure*}

Here, we discuss why both the homogeneous and inhomogeneous prescriptions capture the covariance equally well after applying the simulation-based correction. The right panel of Figure \ref{fig: res_chi2_red_ana_old} suggests that the assumptions of our simulation-based correction method are reasonably met in both cases: that a \textit{rotated} Monte Carlo covariance matrix, $\bm\Sigma_R$ (as defined in Equation \ref{eq: Sigma_R}), is approximately diagonal, and that its diagonal is approximately smooth. We test these predictions in Figure \ref{fig: res_why_does_PSpipe_do_well_coadd_EE}. As expected from Figure \ref{fig: res_chi2_red_ana_old}, the inhomogeneous analytic prescription underestimates the approximate Monte Carlo eigenspectrum by a few percent, as measured by the main diagonal of $\bm\Sigma_R$, and in a smooth fashion easily handled by the simulation-based correction. The homogeneous analytic prescription underestimates the approximate Monte Carlo eigenspectrum by $\sim10-20\%$, although with some prominent features extending as high as $\sim30\%$. Evidently, these features are still easily smoothed by the Gaussian processes in \S\ref{sec: pipeline_sim_correction}.

As discussed in \S\ref{sec: pipeline_sim_correction}, Figure \ref{fig: res_why_does_PSpipe_do_well_coadd_EE} reveals $\bm\Sigma_R$ is not perfectly diagonal within statistical scatter: its off-diagonal blocks ($\beta'\neq\beta$) contain percent-level correlations along their bin-wise diagonals ($b'=b$). As expected, these off-diagonals are slightly more pronounced for the homogeneous matrix prescription, although still small. In either case, they are very smooth, and are thus easily fit by a Gaussian process (we explore the importance of accounting for these small off-diagonals in Appendix \ref{apx: smooth_off_diags}). Most importantly, for either analytic prescription, $\bm\Sigma_R$ contains significantly less structure and dynamic range than the raw covariance or correlation matrices in Figure \ref{fig: res_mc_vs_ana_corrected_mat_coadd_EE}. Moreover, the bin-wise off-diagonal ($b'=b+1$) correlations, which are present at the $\sim5\%$-level in the raw correlation matrix in Figure \ref{fig: res_mc_vs_ana_corrected_mat_coadd_EE}, are absent for $\bm\Sigma_R$, again regardless of the analytic prescription. Therefore, while the core assumptions motivating $\bm\Sigma_R$ are not perfect, they are good enough to enable our simple simulation-based correction scheme. 

Nevertheless, a covariance structure dominated by block-diagonals does not guarantee satisfactory performance when applying any simulation-based matrix correction. For example, a conceptually similar simulation-based correction method to that presented in \S\ref{sec: pipeline_sim_correction} is to assume that the Monte Carlo and analytic matrices share correlation matrices ($\rho_{ij}$) rather than eigenbases. The correction then takes the form of reweighting the analytic matrix diagonal using the Monte Carlo matrix while preserving its correlation matrix. Such a scheme was used in WMAP \citep{Verde2003} and was suggested by \citet{Hamimeche2009}. We also tested this method, where we performed the reweighting of the matrix diagonal using the same Gaussian process smoothing as we developed in \S\ref{sec: pipeline_sim_correction}. We found this simulation-based correction method did not perform as well. Using our inhomogeneous analytic covariance, the corrected matrix resulted in a $\chi^2$ distribution of simulations (as in the right panel of Figure \ref{fig: res_chi2_red_ana_old}) that remained $\sim1.7\%$ too high, while using the homogeneous analytic covariance resulted in a distribution that remained $\sim4.9\%$ too high. Therefore, we conclude that the Monte Carlo and analytic covariance do not share exact correlation structures; rather, the assumption of a shared eigenbasis is more robust. 

While not critical to our covariance validation, for the interested reader we also investigate the effect of point-source holes in the analysis mask (see Figure \ref{fig: data_masks}) on the covariance matrix in Appendix \ref{apx: pt_src_holes}.

\section{Conclusion} \label{sec: conclusion}

We have developed a new power spectrum covariance matrix pipeline for the ACT DR6 data in response to the challenge the data pose to the MASTER covariance framework. While the resulting inhomogenous prescription is similar to that presented in \textit{Planck} \citep{Planck_XV_2013, Planck_XI_2015}, unlike \textit{Planck} the ACT data do not conform to the data model required in MASTER. Through our use of different effective spatial weights for signal and noise, we also depart from existing public covariance implementations. We find that our pipeline yields an analytic covariance matrix for the ACT DR6 data that is accurate to within $\sim3\%$ of a Monte Carlo covariance that incorporates realistic properties of the ACT data. While we observe $\sim16\%$-level differences between an analogous homogeneous matrix and the Monte Carlo matrix, both analytic matrices achieve sub-percent agreement with the Monte Carlo covariance after applying new a simulation-based correction method. In comparison, a common correction method assuming that the analytic correlation matrix is accurate, while only correcting the covariance diagonal, results in biases as large as $\sim5\%$. Our central result is that a semi-analytic covariance using our simulation-based correction and either the inhomogeneous or homogeneous prescriptions is well-suited for use in the ACT DR6 likelihood.

This work has clarified the mechanisms behind the $\sim16\%$-level discrepancies between the Monte Carlo and the homogeneous analytic covariance matrix. Accepting simulation-based corrections of that magnitude, without understanding what is driving them, may have limited our confidence in the final covariance matrix. Instead, we understand most of the salient features of the simulations at the covariance level: by making a set of \textit{a priori} well-motivated changes to the pipeline, summarized in \S\ref{sec: pipeline_analytic_comparison}, we achieve a discrepancy between the analytic and Monte Carlo covariance of less than $3\%$. This gives us sufficient confidence in the simulations to condition the homogeneous analytic covariance, even for corrections as large as $\sim16\%$. 

This work has implications for future large-aperture CMB experiments, such as the Simons Observatory (SO). A change to the map-based noise properties, or a change in power spectrum bin size, could result in a noticeable difference in performance between the homogeneous and inhomogeneous prescriptions, even after attempts at matrix conditioning. The homogeneous prescription is also more reliant on a well-converged Monte Carlo covariance, so this new prescription may allow for a smaller simulation ensemble. Between this new analytic prescription and continued simulation development, we anticipate this power spectrum covariance matrix will continue to meet requirements for next-generation CMB science programs.

\section*{Acknowledgments}
ZA and JD acknowledge support from NSF grant AST-2108126. This work was supported by a grant from the Simons Foundation (CCA 918271, PBL). The Flatiron Institute is supported by the Simons Foundation. CS acknowledges support from the Agencia Nacional de Investigaci\'on y Desarrollo (ANID) through Basal project FB210003. KM acknowledges support from the National Research Foundation of South Africa. EC and HJ acknowledge support from the Horizon 2020 ERC Starting Grant (Grant agreement No 849169); SG and EC also acknowledge support from STFC and UKRI (grant numbers ST/W002892/1 and ST/X006360/1). Support for ACT was through the U.S.~National Science Foundation through awards AST-0408698, AST-0965625, and AST-1440226 for the ACT project, as well as awards PHY-0355328, PHY-0855887 and PHY-1214379. Funding was also provided by Princeton University, the University of Pennsylvania, and a Canada Foundation for Innovation (CFI) award to UBC. ACT operated in the Parque Astron\'omico Atacama in northern Chile under the auspices of the Agencia Nacional de Investigaci\'on y Desarrollo (ANID). The development of multichroic detectors and lenses was supported by NASA grants NNX13AE56G and NNX14AB58G. Detector research at NIST was supported by the NIST Innovations in Measurement Science program. Computing for ACT was performed using the Princeton Research Computing resources at Princeton University, the National Energy Research Scientific Computing Center (NERSC), and the Niagara supercomputer at the SciNet HPC Consortium. SciNet is funded by the CFI under the auspices of Compute Canada, the Government of Ontario, the Ontario Research Fund–Research Excellence, and the University of Toronto. We thank the Republic of Chile for hosting ACT in the northern Atacama, and the local indigenous Licanantay communities whom we follow in observing and learning from the night sky.

Some of the results in this paper have been derived using the \texttt{healpy} and \texttt{HEALPix} package \citep{healpy, HEALPix}. Other software used in this paper include the following: \software{\texttt{astropy} \citep{astropy:2013, astropy:2018, astropy:2022}, \texttt{cython} \citep{cython}, \texttt{ducc} (\url{https://gitlab.mpcdf.mpg.de/mtr/ducc}), \texttt{h5py} \citep{h5py}, \texttt{libsharp} \citep{libsharp}, \texttt{matplotlib} \citep{matplotlib}, \texttt{mnms} \citep{mnms}, \texttt{numba} \citep{numba}, \texttt{numpy} \citep{numpy}, \texttt{pyfftw} \citep{pyfftw}, \texttt{pyyaml} (\url{https://pyyaml.org/}), and \texttt{scipy} \citep{scipy}.} 

\bibliography{main}
\bibliographystyle{aasjournal}

\appendix

\section{More Details on Data Model} \label{apx: sims}

We presented the ACT data model in \S\ref{sec: formalism_full_data_model}; for brevity, we omitted some pieces that we document here. These pieces do not fundamentally change the data description, but do complicate our map-level simulation somewhat beyond what is prescribed in Equation \ref{eq: full_data_model}. The simulations in \S\ref{sec: pipeline_sims} follow the additional steps of this section.

The beam-convolved sky signal defined in Equation \ref{eq: signal_def} does not exactly correspond to the signal in the ACT maps. Instead, optical and readout inefficiencies result in an overall calibration error in the maps, and the map digitization --- or ``mapmaking" --- introduces a pixel window function:
\begin{align} \label{eq: apx_signal_def}
    \mathbf s = \mathbf A \mathbf F^{\dag} \mathbf X_p \mathbf F \mathbf Y \mathbf B \mathbf{S}^{\frac{1}{2}} \bm\eta_s,
\end{align}
where $\mathbf F^{\dag} \mathbf X_p \mathbf F$ applies a pixel window function matrix \mbf{X_p}, which is diagonal Fourier space, to the maps, and \mbf{A} is a diagonal matrix containing the observational efficiency factor for each array and polarization. We account for the map noise ``as is,", in that Equation \ref{eq: noise_def} models the noise in the map products themselves, and so it is arbitrary whether we include the calibration or pixel window in the noise model as long as we are consistent. We opt to exclude them.  

As discussed in \S\ref{sec: formalism_full_data_model}, after the ACT maps are made, they are further processed as part of the power spectrum estimation pipeline (the ACT DR6 power spectrum paper will provide a more detailed explanation). In \S\ref{sec: formalism_full_data_model}, we omitted two steps in that processing. Firstly, measurement of the ACT observational efficiency in both temperature and polarization is made through spectrum-level comparisons against \textit{Planck}. These corrections are defined and applied at the map level. In terms of our data model from \S\ref{sec: formalism_full_data_model}, this looks like constructing \mbf{A^{-1}} and applying it to \mbf{m}: $\mathbf A^{-1} \mathbf m$. Next, the pixel window function is divided out in Fourier space. In fact, because both the pixel window function and the Fourier filter are diagonal in Fourier space, both steps are performed simultaneously. Importantly, since these operations are slightly non-local in map space, prior to performing them the maps are masked with a broad, apodized sky mask which eliminates bright galactic regions and noisy pixels near the edge of the ACT footprint. Altogether, this applies the following matrix to our calibrated map model: $\mathbf F^{\dag} \mathbf X_f \mathbf X_p^{-1} \mathbf F \mathbf W_k$. Here, \mbf{W_k} is the sky mask applied before the Fourier operations, \mbf{X_p} is the pixel window function, and \mbf{X_f} is the pickup filter in Fourier space. Together with applying the analysis mask (\mbf{W}) as discussed in \S\ref{sec: formalism_full_data_model}, we have the following data model of the processed ACT maps (again, in the case of the ``tiled" noise model):
\begin{equation} \label{eq: sim_def}
    \begin{aligned}
        \mathbf m_i &\rightarrow \mathbf W \mathbf F^{\dag} \mathbf X_f \mathbf X_p^{-1} \mathbf F \mathbf W_k \mathbf A^{-1} \mathbf m_i \\
        &= \mathbf W \mathbf F^{\dag} \mathbf X_f \mathbf X_p^{-1} \mathbf F \mathbf W_k \mathbf F^{\dag} \mathbf X_p \mathbf F \mathbf Y \mathbf B \mathbf{S}^{\frac{1}{2}} \bm\eta_s \\
        &+ \mathbf W \mathbf F^{\dag} \mathbf X_f \mathbf X_p^{-1} \mathbf F \mathbf W_k \mathbf A^{-1} \bm\sigma_i \mathbf Y \mathbf N_i^{\frac{1}{2}} \mathbf Y^{\dag}\bm\Omega \mathcal{T}^{\dag} \mathbf F^{\dag} \mathcal{N}_i^{\frac{1}{2}}\bm\eta_{n,i}.
    \end{aligned}
\end{equation}
This is the full forward model for the DR6 simulated maps discussed in \S\ref{sec: pipeline_sims}.

In practice, Equation \ref{eq: sim_def} is too cumbersome to work with for the analytic covariance matrix, so we make further simplifying assumptions. We discussed one in \S\ref{sec: formalism_approx_data_model}: the approximate noise model (Equation \ref{eq: approx_noise_def}). Furthermore, since the isotropic noise power spectrum is measured from the data (see \S\ref{sec: pipeline_analytic_noise}), we can absorb the map calibration operation, \mbf{A^{-1}}, into it. Then, for both the signal and noise terms, we assume that after applying the analysis mask, \mbf{W}, edge effects from the Fourier operations on the \mbf{W_k}-masked data are indeed negligible, such that we can eliminate \mbf{W_k} from Equation \ref{eq: full_data_model}. Finally, since the pixel window function (and its inverse), \mbf{X_p}, is almost isotropic, and the noise is already highly anisotropic, we also absorb \mbf{X_p^{-1}} into the isotropic noise power spectrum of the approximate noise model. In other words, neglecting the anisotropy of \mbf{X_p^{-1}} is a negligible addition to the data model error already incurred by the approximate noise model. Cancelling terms leads to the following simplified data model of the processed maps: 
\begin{align}  \label{eq: sim_def2}
    \mathbf m_i = \mathbf W (\mathbf F^{\dag} \mathbf X_f \mathbf F) \mathbf Y \mathbf B \mathbf{S}^{\frac{1}{2}} \bm\eta_s + \mathbf W (\mathbf F^{\dag} \mathbf X_f \mathbf F) \bm\sigma_i \bm\Omega^{\frac{1}{2}} \mathbf Y \mathbf N_i^{\frac{1}{2}} \bm\eta_{n,i},
\end{align}
where \mbf{N_i} is measured \textit{after} the map processing, and we have placed in parentheses the Fourier-space pickup filter (the $\mathbf F^{\dag} \mathbf X_f \mathbf F$ operator). Replacing the Fourier-space pickup filter with an effective transfer function in harmonic space, as described in \S\ref{sec: formalism_approx_data_model}, yields Equation \ref{eq: approx_data_model}.

We also show the origin of the $\bm\Omega^{\frac{1}{2}}$ factor in the effective noise weight when going from Equation \ref{eq: almost_approx_noise_def} to Equation \ref{eq: approx_noise_def}. The factor appears in the literature as early as \citet{Efstathiou2004}, and recurs throughout the \textit{Planck} covariance matrices, but its whereabouts have not been explicitly discussed. A field that follows the model of Equation \ref{eq: almost_approx_noise_def} has zero mean and is Gaussian, so its statistics are fully specified by its covariance. This is given by:
\begin{equation} \label{eq: sqrt_pixar_trick}
    \begin{aligned}
        \langle\mathbf n_i\mathbf n_i^\dag\rangle &= \bm\sigma_i \mathbf Y \mathbf N_i^{\frac{1}{2}} \mathbf Y^{\dag}\bm\Omega \langle\bm\eta_{n,i}\bm\eta_{n,i}^\dag\rangle\bm\Omega\mathbf Y \mathbf N_i^{\frac{T}{2}}\mathbf Y^\dag \bm\sigma_i \\
        &= \bm\sigma_i \mathbf Y \mathbf N_i^{\frac{1}{2}} \mathbf Y^{\dag}\bm\Omega^2\mathbf Y \mathbf N_i^{\frac{T}{2}}\mathbf Y^\dag \bm\sigma_i \\
        &\approx \bm\sigma_i\bm\Omega^{\frac{1}{2}} \mathbf Y \mathbf N_i^{\frac{1}{2}} (\mathbf Y^{\dag}\bm\Omega\mathbf Y) \mathbf N_i^{\frac{T}{2}}\mathbf Y^\dag \bm\Omega^{\frac{1}{2}}\bm\sigma_i \\
        &= \bm\sigma_i\bm\Omega^{\frac{1}{2}} \mathbf Y \mathbf N_i\mathbf Y^\dag \bm\Omega^{\frac{1}{2}}\bm\sigma_i,
    \end{aligned}
\end{equation}
where the following definitions and properties were used: that $\bm\eta_{n,i}$ is a white-noise vector, that $\bm\Omega^{\frac{1}{2}}$ is a smooth map that nearly commutes with the harmonic operations in terms like $\mathbf Y \mathbf N_i^{\frac{1}{2}} \mathbf Y^{\dag}$, and that $\mathbf Y^{\dag}\bm\Omega\mathbf Y$ is the identity. The covariance in the last line of Equation \ref{eq: sqrt_pixar_trick} is the same as that for a field following the model of Equation \ref{eq: approx_noise_def}. Therefore, fields following Equations \ref{eq: almost_approx_noise_def} and \ref{eq: approx_noise_def} are nearly identically distributed.

\section{Analytical Pseudospectrum Covariance Matrices} \label {apx: covmat}

In this section, we derive Equation \ref{eq: general_pseudocov} and give our expressions for the coupling ``spin," $\beta$. We also motivate our use of an isotropic transfer function to approximately model the Fourier-space filter, as described in \S\ref{sec: pipeline_analytic_fourier}, and the steps to measure each of its components.

\subsection{Covariance Matrices} \label{apx: covmat_core}
 
The structure of the pseudospectrum covariance matrix term in Equation \ref{eq: general_pseudocov} is similar to that of \citet{Planck_XI_2015} and \citet{INKA}, but here we provide an explicit derivation for reference. We work in the temperature-only case, but the derivation is analogous for polarization \citep[see e.g.,][for the appropriate substitutions]{Challinor2004,Brown2005,Couchot2017,Camphuis2022}.

We start by defining a few mathematical tools. First, we can write Equation \ref{eq: scalar_MASTER_model} in harmonic space as \citep[see e.g.,][]{MASTER}:
\begin{align} \label{eq: pseudoharmonic}
    \tilde{a}_{\ell m} = \sum_{\ell'm'}K_{\ell m,\ell'm'}(w)a_{\ell'm'},
\end{align}
where
\begin{align}
    K_{\ell m,\ell'm'}(w) \equiv \int dxY_{\ell'm'}(x)w(x)Y_{\ell m}^*(x).
\end{align}
The $K$ matrices have the following property \citep[see e.g.,][]{Couchot2017}:
\begin{align} \label{eq: K_inner_product}
    K_{\ell_1 m_1, \ell_2 m_2}(yz) = \sum_{\ell_3 m_3}K_{\ell_1 m_1, \ell_3 m_3}(y)K_{\ell_3 m_3, \ell_2 m_2}(z) = \sum_{\ell_3 m_3}K_{\ell_1 m_1, \ell_3 m_3}(y)K_{\ell_2 m_2, \ell_3 m_3}^*(z),
\end{align}
where $yz$ is the product of the two masks $y$ and $z$ in map space. Finally, we need the definition of the coupling matrices, $\Xi_{\ell\ell'}(y, z)$, from Equation \ref{eq: 2pt_MASTER}:
\begin{align} \label{eq: xi_def}
    \Xi_{\ell\ell'}(y, z) \equiv \frac{1}{2\ell+1}\frac{1}{2\ell'+1}\sum_{m m'}K_{\ell m,\ell'm'}(y)K_{\ell m,\ell'm'}^*(z).
\end{align}
As noted in \citet{MASTER}, Equation \ref{eq: xi_def} can be calculated in terms of the cross-power spectrum of the masks $y$ and $z$ and the Wigner 3-j symbols in $\mathcal{O}(\ell_{max}^3)$-time. 

Next, consider four instances of an isotropic field, $a$, and mask, $w$, that follow the model of Equation \ref{eq: vector_MASTER_model}. We label each instance $i$, $j$, $p$, and $q$. The fields are fully specified by their cross-power spectra, $C_\ell^{yz}$, where $yz$ is a pair of labels:
\begin{align} \label{eq: isotropic}
    \langle a_{\ell m}^y a_{\ell'm'}^z\rangle_s = C_\ell^{yz}\delta_{\ell\ell'}\delta_{mm'}.
\end{align}
The pseudospectrum estimator for a pair of these fields (Equation \ref{eq: 2pt_MASTER}) is $\hat{\tilde{C}}_\ell^{yz}$, and the covariance of $i$, $j$, $p$, and $q$ is:
\begin{equation} \label{eq: apx_cov}
    \begin{aligned}
        \tilde{\Sigma}_{\ell\ell'}^{i,j,p,q} &\equiv \langle\hat{\tilde{C}}_\ell^{ij} \hat{\tilde{C}}_{\ell'}^{pq}\rangle - \langle\hat{\tilde{C}}_\ell^{ij}\rangle\langle\hat{\tilde{C}}_\ell^{pq}\rangle \\
        &= \frac{1}{2\ell+1}\frac{1}{2\ell'+1}\sum_{m,m'}\langle \tilde a_{\ell m}^i \tilde a_{\ell m}^{j*} \tilde a_{\ell'm'}^p \tilde a_{\ell'm'}^{q*}\rangle - \langle\tilde a_{\ell m}^i \tilde a_{\ell m}^{j*}\rangle\langle\tilde a_{\ell'm'}^p \tilde a_{\ell'm'}^{q*}\rangle.
    \end{aligned}
\end{equation}
Because $\tilde a$ is Gaussian, we expand the product of four fields using Wick's theorem. Also using the reality of the fields, Equation \ref{eq: apx_cov} becomes:
\begin{align} \label{eq: after_Wick}
    \tilde{\Sigma}_{\ell\ell'}^{i,j,p,q} = \frac{1}{2\ell+1}\frac{1}{2\ell'+1}\sum_{m,m'}\langle\tilde a_{\ell m}^i \tilde a_{\ell'm'}^{p*}\rangle\langle\tilde a_{\ell m}^{j*} \tilde a_{\ell'm'}^q\rangle + \langle\tilde a_{\ell m}^i \tilde a_{\ell'm'}^{q*}\rangle\langle\tilde a_{\ell m}^{j*} \tilde a_{\ell'm'}^p\rangle.
\end{align}
Combining with Equations \ref{eq: pseudoharmonic} and \ref{eq: isotropic}, we get:
\begin{equation} \label{eq: before_NKA}
    \begin{aligned} 
        \tilde{\Sigma}_{\ell\ell'}^{i,j,p,q} = \frac{1}{2\ell+1}\frac{1}{2\ell'+1}\sum_{m,m'}&\sum_{\ell_1 m_1}K_{\ell m,\ell_1 m_1}(w^i)K_{\ell'm',\ell_1 m_1}^*(w^p)C_{\ell_1}^{ip}\sum_{\ell_3 m_3}K_{\ell m,\ell_3 m_3}^*(w^j)K_{\ell'm',\ell_3 m_3}(w^q)C_{\ell_3}^{jq} + \\ 
        + &\sum_{\ell_1 m_1}K_{\ell m,\ell_1 m_1}(w^i)K_{\ell'm',\ell_1 m_1}^*(w^q)C_{\ell_1}^{iq}\sum_{\ell_3 m_3}K_{\ell m,\ell_3 m_3}^*(w^j)K_{\ell'm',\ell_3 m_3}(w^p)C_{\ell_3}^{jp}.
    \end{aligned}
\end{equation}
Each term like $\sum_{\ell_1 m_1}K_{\ell m,\ell_1 m_1}(w^i)K_{\ell'm',\ell_1 m_1}^*(w^p)C_{\ell_1}^{ip}$ is difficult to compute directly, but if we assume the NKA, then we can move $C_{\ell_1}^{ip}$ out of the sum over $\ell_1$, yielding:
\begin{align} \label{eq: the_NKA}
    \sum_{\ell_1 m_1}K_{\ell m,\ell_1 m_1}(w^i)K_{\ell'm',\ell_1 m_1}^*(w^p)C_{\ell_1}^{ip} \rightarrow C_{(\ell,\ell')}^{ip}\sum_{\ell_1 m_1}K_{\ell m,\ell_1 m_1}(w^i)K_{\ell'm',\ell_1 m_1}^*(w^p),
\end{align}
where $(\ell,\ell')$ denotes some symmetric function of $\ell$ and $\ell'$ \citep{Efstathiou2004}. Finally, combining Equations \ref{eq: K_inner_product}, \ref{eq: xi_def}, \ref{eq: before_NKA}, and \ref{eq: the_NKA} yields:
\begin{align} \label{eq: after_NKA}
    \tilde{\Sigma}_{\ell\ell'}^{i,j,p,q} = C_{(\ell,\ell')}^{ip}C_{(\ell,\ell')}^{jq}\Xi_{\ell\ell'}(w^iw^p,w^jw^q) + C_{(\ell,\ell')}^{iq}C_{(\ell,\ell')}^{jp}\Xi_{\ell\ell'}(w^iw^q,w^jw^p).
\end{align}

In the case of Equation \ref{eq: general_data_model}, we have additive signal (fields $s$, masks $u$, and power spectra $S$) and noise (fields $n$, masks $v$, and power spectra $N$) that are uncorrelated. Substituting Equation \ref{eq: general_data_model} for $\tilde a$ in Equation \ref{eq: after_Wick}, and expanding into signal and noise, would yield 16 $(ip,jq)$ signal-noise cross terms and 16 $(iq,jp)$ signal-noise cross terms, but assuming the signal and noise are uncorrelated reduces this to four of each term, in particular:
\begin{equation} \label{eq: apx_cov2}
    \begin{aligned}
        \tilde{\Sigma}_{\ell\ell'}^{i,j,p,q} &= S_{(\ell,\ell')}^{ip}S_{(\ell,\ell')}^{jq}\Xi_{\ell\ell'}(u^i u^p, u^j u^q) 
        + S_{(\ell,\ell')}^{ip}N_{(\ell,\ell')}^{jq}\Xi_{\ell\ell'}(u^i u^p, v^j v^q)  \\
        &+ N_{(\ell,\ell')}^{ip}S_{(\ell,\ell')}^{jq}\Xi_{\ell\ell'}(v^i v^p, u^j u^q)  
        + N_{(\ell,\ell')}^{ip}N_{(\ell,\ell')}^{jq}\Xi_{\ell\ell'}(v^i v^p, v^j v^q)  \\
        &+ S_{(\ell,\ell')}^{iq}S_{(\ell,\ell')}^{jp}\Xi_{\ell\ell'}(u^i u^q, u^j u^p)  
        + S_{(\ell,\ell')}^{iq}N_{(\ell,\ell')}^{jp}\Xi_{\ell\ell'}(u^i u^q, v^j v^p)  \\
        &+ N_{(\ell,\ell')}^{iq}S_{(\ell,\ell')}^{jp}\Xi_{\ell\ell'}(v^i v^q, u^j u^p)  
        + N_{(\ell,\ell')}^{iq}N_{(\ell,\ell')}^{jp}\Xi_{\ell\ell'}(v^i v^q, v^j v^p),
    \end{aligned}
\end{equation}
which is equivalent to Equation \ref{eq: general_pseudocov}. 

To fully specify our covariance matrix prescription in Equation \ref{eq: general_pseudocov}, we provide the $\beta$ mapping from field polarizations to coupling spins in Table \ref{tab: spin_mapping}. This follows from the INKA, as well as the assumption that mask gradients can be neglected \citep{Couchot2017}, such that ``$--$" couplings are small compared to ``++" couplings.

\begin{table}
    \centering
    \begin{tabular}{c|l}
         $\beta(AB,CD)$ & $AB, CD$ \\
         \hline 
         \hline
         \multirow{3}{*}{00} & TT, TT \\
         & TT, TP + three permutations \\
         & TP, TP + three permutations \\
         \hline
         \multirow{2}{*}{0+} & TT, PP + one permutation \\
         & TP, PP + 3 permutations \\
         \hline
         ++ & PP, PP \\
         \hline 
         $--$ & Not used
    \end{tabular}
    \caption{Coupling spins as a function of input field polarizations. $P$ refers to either $E$ or $B$. Note, there are 16 possible polarization permutations, and in no case do we use the $--$ coupling.}
    \label{tab: spin_mapping}
\end{table}

\subsection{More Details on Fourier Filter} \label{apx: covmat_kspace}

In this section, we first examine the effect of a Fourier-space filter analytically by considering it in harmonic space, and then discuss how we construct our approximation for the Fourier-space filter described in \S\ref{sec: pipeline_analytic_fourier}.

A filter applied to fields in Fourier (or harmonic) space has analogous effects on the power spectrum and covariance matrix as a mask applied in map space. For example, the different manifestation of the mask at the spectrum (two-point) level and covariance (four-point) level results in the factor of the mask sky fraction, $f_{sky}$, in the denominator of the leading-order power spectrum covariance matrix \citep{Knox1995}. To derive a similar effect for a filter, first consider a filter applied to fields in harmonic space in the absence of any mask. The field has the following data model:
\begin{align}
    \tilde a_{\ell m} = f_{\ell m} C_\ell^{\frac{1}{2}} \eta_{\ell m},
\end{align}
where the filter $f_{\ell m}$ can be anisotropic ($m$-dependent). In expectation, the ``pseudospectrum" of $\tilde a$ is then related to the power spectrum $C_\ell$ as:
\begin{align} \label{eq: filt_step_1}
    \langle\hat{\tilde{C}}_{\ell}\rangle = \frac{1}{2\ell+1} \sum_{m=-\ell}^{\ell} \langle\tilde a_{\ell m} \tilde a_{\ell m}^*\rangle = \frac{1}{2\ell+1} \sum_{m=-\ell}^{\ell} f_{\ell m}^2 C_\ell \equiv t_{\ell,2pt} C_\ell,
\end{align}
where we have defined the two-point isotropic transfer function:
\begin{align}
    t_{\ell,2pt} \equiv \frac{1}{2\ell+1} \sum_{m=-\ell}^{\ell} f_{\ell m}^2.
\end{align}
The covariance of the pseudospectrum is given by Equation \ref{eq: after_Wick}:
\begin{equation}
    \begin{aligned}
        \tilde{\Sigma}_{\ell\ell'} \equiv \frac{2}{(2\ell+1)(2\ell'+1)}\sum_{m,m'}\langle|\tilde a_{\ell m} \tilde a_{\ell'm'}^*|^2\rangle &= \frac{2}{(2\ell+1)(2\ell'+1)}\sum_{m,m'}|f_{\ell m}f_{\ell'm'}C_\ell^{\frac{1}{2}}C_{\ell'}^{\frac{1}{2}}\langle\delta_{\ell\ell'}\delta_{mm'}\rangle|^2 \\
        &= \frac{2\delta_{\ell\ell'}}{(2\ell+1)^2}\sum_m f_{\ell m}^4 C_\ell^2 \\
        &\equiv \frac{2t_{\ell,4pt}C_\ell^2\delta_{\ell\ell'}}{2\ell+1}, 
    \end{aligned}
\end{equation}
where we have defined the four-point isotropic transfer function:
\begin{align}
    t_{\ell,4pt} \equiv \frac{1}{2\ell+1} \sum_{m=-\ell}^{\ell} f_{\ell m}^4.
\end{align}
Then the covariance matrix of the \textit{power spectrum} is given by:
\begin{align}
    \Sigma_{\ell\ell'} = \frac{1}{t_{\ell,2pt}t_{\ell',2pt}}\tilde{\Sigma}_{\ell\ell'} = \frac{2t_{\ell,4pt}C_\ell^2\delta_{\ell\ell'}}{t_{\ell,2pt}^2(2\ell+1)}.
\end{align}
The filter factor $t_{\ell,4pt}/t_{\ell,2pt}^2$ has non-trivial behavior depending on the character of $f_{\ell m}$. If $f_{\ell m}$ is actually isotropic --- with no $m$-dependence, like a beam --- then $t_{\ell,4pt}/t_{\ell,2pt}^2=1$. If $f_{\ell m}$ is binary --- equal only to 0 or 1 --- then $t_{\ell,4pt}=t_{\ell,2pt}\equiv t_\ell$ and $t_{\ell,4pt}/t_{\ell,2pt}^2=1/t_\ell$, where $t_\ell\equiv 1/(2\ell+1)\sum_m f_{\ell m}$ is the average value of $f_{\ell m}$ over $m$. If $f_{\ell m}$ has any anisotropy, then it is less than 1. In this case, we see an analogy to a mask: the covariance matrix increases as $1/t_\ell$.

When there is a mask in addition to a harmonic-space filter, these expressions become inexact. The pseudospectrum is then related to the power spectrum as:
\begin{align}
    \langle\hat{\tilde{C}}_{\ell}\rangle = \frac{1}{2\ell+1}\sum_{\ell'}C_{\ell'}\sum_{mm'}K_{\ell m,\ell'm'}(w)K_{\ell m,\ell'm'}^*(w)f_{\ell'm'}^2.
\end{align}
Ordinarily, at this point, the MASTER formalism would use Equation \ref{eq: xi_def}, but if $f_{\ell m}$ is anisotropic, this simplification is no longer possible. To recover the MASTER formalism, we approximate $f_{\ell m}^2$ as being equal to its isotropic average, raised to some exponent, $\alpha_{2pt}$:
\begin{equation} \label{eq: 2pt_ansatz}
    \sum_{mm'}K_{\ell m,\ell'm'}(w)K_{\ell m,\ell'm'}^*(w)f_{\ell'm'}^2 \rightarrow t_\ell^{\alpha_{2pt}}\sum_{mm'}K_{\ell m,\ell'm'}(w)K_{\ell m,\ell'm'}^*(w),
\end{equation}
where:
\begin{align} \label{eq: fl_2pt_def}
    t_\ell\equiv\frac{1}{2l+1}\sum_m f_{\ell m}^2.
\end{align}
To be sure, the substitution in Equation \ref{eq: 2pt_ansatz} is an ansatz meant to approximately capture the actual expression. The ansatz assumes the effect of the harmonic-space filter is entirely imparted on the underlying power spectrum of the filtered field, with no effect on the mode-coupling matrix $M_{\ell\ell'}$. Overall, we find this ansatz works quite well, and is independent of whether Equation \ref{eq: fl_2pt_def} is binned, indicating that there are not significant changes to the mode-coupling structure. We then have the following for the pseudospectrum:
\begin{align} \label{eq: 2pt_filter_fwd_model}
    \langle\hat{\tilde{C}}_{\ell}\rangle = \sum_{\ell'}M_{\ell\ell'}t_\ell^{\alpha_{2pt}}C_{\ell'},
\end{align}
as discussed in \S\ref{sec: pipeline_analytic_fourier}. 

For the pseudospectrum covariance, we go back to Equation \ref{eq: the_NKA}, which in the presence of a harmonic-space filter becomes:
\begin{align} \label{eq: 4pt_ansatz}
    \sum_{\ell_1 m_1}K_{\ell m,\ell_1 m_1}(w)K_{\ell'm',\ell_1 m_1}^*(w)f_{\ell_1 m_1}^2 C_{\ell_1} \rightarrow (t^{\alpha_{4pt}}C)_{(\ell,\ell')}\sum_{\ell_1 m_1}K_{\ell m,\ell_1 m_1}(w)K_{\ell'm',\ell_1 m_1}^*(w),
\end{align}
where we have made a similar ansatz as Equation \ref{eq: 2pt_ansatz}, but as it involves a different sum (over $\ell$ and $m$, not just over $m$), we assign a different exponent to $t_\ell$ --- $\alpha_{4pt}$ --- and we still use the NKA: $(t^{\alpha_{4pt}}C)_{(\ell,\ell')}$ denotes a symmetric function of $t_\ell^{\alpha_{4pt}}C_\ell$ and $t_{\ell'}^{\alpha_{4pt}}C_{\ell'}$. The pseudospectrum covariance matrix (Equation \ref{eq: after_NKA}) then becomes:
\begin{align} \label{eq: 4pt_filter_pseudo_fwd_model}
    \tilde{\Sigma}_{\ell\ell'} = 2(t^{\alpha_{4pt}}C)_{(\ell,\ell')}^2\Xi_{\ell\ell'}(w^2,w^2).
\end{align}
We have made the same assumption that the effect of the harmonic-space filter is only on the underlying power spectrum and not on the coupling matrix. We find that this ansatz performs reasonably well, but unlike Equation \ref{eq: 2pt_filter_fwd_model}, we find the binning does matter, and also find the preferred value of $\alpha_{4pt}$ changes when using an expression for the binned \textit{power spectrum} covariance matrix:
\begin{align} \label{eq: 4pt_filter_power_fwd_model}
    \Sigma_{bb'} = \sum_{\ell}Q_{b\ell}\sum_{\ell'}Q_{b'\ell'}\tilde{\Sigma}_{\ell\ell'} = 2\sum_{\ell}Q_{b\ell}\sum_{\ell'}Q_{b'\ell'}(t^{\alpha_{4pt}}C)_{(\ell,\ell')}^2\Xi_{\ell\ell'}(w^2,w^2),
\end{align}
where, unlike Equation \ref{eq: pseudo2spec_2pt}, here we use a simplified, diagonal \mbf{U} matrix with diagonal equal to $1/t_\ell^{\alpha_{2pt}}$, following Equation \ref{eq: 2pt_filter_fwd_model}. These findings indicate that this ansatz is not optimal, but given our simulation-based correction, it is sufficient.

\begin{figure}
    \centering
    \includegraphics[width=0.85\textwidth]{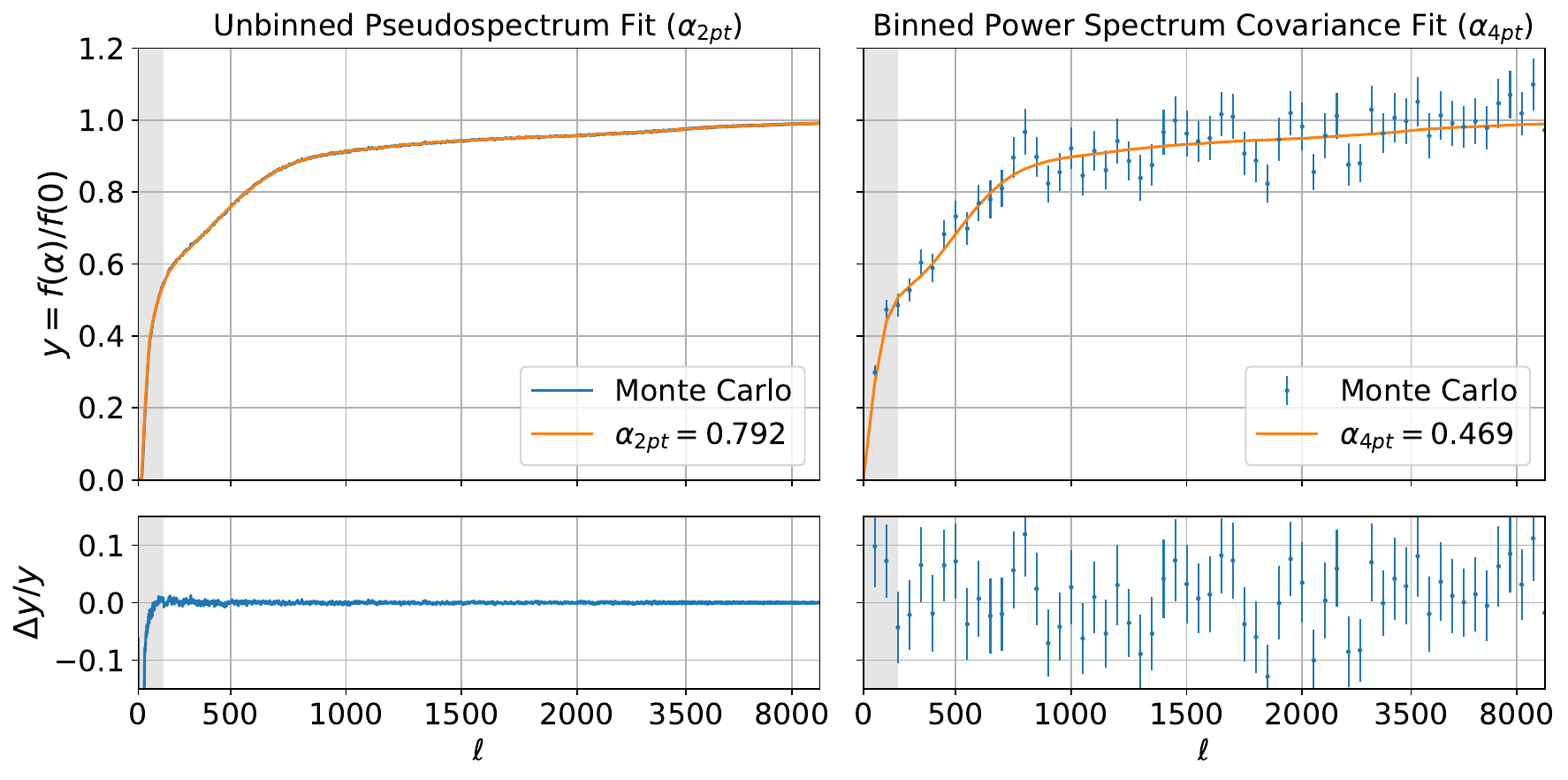}
    \caption{\textit{Left:} Fitting the model of Equation \ref{eq: 2pt_filter_fwd_model} to simulated pseudospectra by optimizing $\alpha_{2pt}$. \textit{Right:} Fitting the model of Equation \ref{eq: 4pt_filter_power_fwd_model} to simulated Monte Carlo covariance diagonals by optimizing $\alpha_{4pt}$. In both cases, the fits are performed (and plotted) after normalizing the model by the case of $\alpha=0$. The model is denoted by $f$, and the normalized model is denoted by $y$. Bottom panels give the residuals relative to the fitted model. The fit only uses scales where $t_\ell>0.5$; the excluded large-scales are shaded in grey.}
    \label{fig: filter_fits}
\end{figure}

In reality, our filter lives in Fourier space, not harmonic space, but we follow the setup of the preceding discussion regardless in building a MASTER-compatible filter approximation. Since our analysis includes a mask, we fit for each quantity of interest --- $t_\ell$, $\alpha_{2pt}$, and $\alpha_{4pt}$ --- using simulations. We first draw 50 full-sky, white-noise simulations, filter them in Fourier space, and measure the transfer function template, $t_\ell$, directly via Equation \ref{eq: filt_step_1}. We have a prior assumption that $t_\ell$ should be smooth, so we can low-pass filter the noisy Monte Carlo estimate of $t_\ell$ to get a better measurement. We use a Savitzky-Golay filter (implemented in \texttt{scipy.signal.savgol\_filter}) with a window length of 100 and a polynomial order of 4. As the basis of our fit for $\alpha_{2pt}$ and $\alpha_{4pt}$, we next draw an ensemble of 500 simulations that include masking. These simulations are drawn according to the following data model for a single scalar field:
\begin{align} \label{eq: filter_sims}
    \mathbf a = \mathbf W (\mathbf F^{\dag} \mathbf X_f \mathbf F) \mathbf W_k \mathbf Y \mathbf B \mathbf{C}^{\frac{1}{2}} \bm\eta,
\end{align}
which, compared to Equation \ref{eq: sim_def}, omits all anisotropic features other than the Fourier filter (i.e., instrinsic noise anisotropies and the pixel window), but is otherwise realistic. As mentioned in \S\ref{sec: pipeline_analytic_fourier}, we use a single representative sky mask \mbf{W} (and pre-filter mask \mbf{W_k}), and only one representative power spectrum \mbf{C} for each of temperature and polarization. The sky mask \mbf{W} is the average of all effective-noise-weight masks $v_{I_i}^X$ (defined in \S\ref{sec: pipeline_analytic}) in DR6; the pre-filter mask \mbf{W_k} is the average of all pre-filter masks (one for each array) in DR6; the power spectra are defined by:
\begin{equation} \label{eq: mock_ps}
    C_\ell = 
    \begin{cases}
        (\ell/\ell_{knee})^p + 1 & \ell \geq \ell_{cap} \\
        (\ell_{cap}/\ell_{knee})^p + 1 & \ell < \ell_{cap} \\
    \end{cases},
\end{equation}
where for temperature $\ell_{knee}=3,000$, $\ell_{cap}=300$, and $p=-4$, and for polarization $\ell_{knee}=300$, $\ell_{cap}=100$, and $p=-4$. These spectra were chosen because they roughly resemble the observed noise power spectra without the Fourier filter. We measure the pseudospectra and Monte Carlo covariance matrix of the simulation ensemble. To fit for $\alpha_{2pt}$, we use our measured $t_\ell$ template and optimize Equation \ref{eq: 2pt_filter_fwd_model}, where the mode-coupling matrix is built using the representative sky mask, against the mean and scatter of the simulated pseudospectra. To fit for $\alpha_{4pt}$, we use $t_\ell$ and optimize Equation \ref{eq: 4pt_filter_power_fwd_model} against the mean and scatter of the simulated Monte Carlo covariance matrix diagonals. Within Equation \ref{eq: 4pt_filter_power_fwd_model}, we use the INKA approximation with arithmetic symmetry (defined in \S\ref{sec: pipeline_analytic}), where the mode-coupling matrix and coupling are both built using the representative sky mask. Results of those fits for the polarization case are shown in Figure \ref{fig: filter_fits}. The fit for the pseudospectra ($\alpha_{2pt}=0.792$) is excellent, with errors at at the sub-percent level. The fit for the binned power spectrum covariance ($\alpha_{4pt}=0.469$) appears consistent given the noisier Monte Carlo estimates. When the final filter transfer functions --- $t_\ell^{\alpha_{2pt}}$ and $t_\ell^{\alpha_{4pt}}$ --- are applied to the fiducial signal and noise spectra in \S\ref{sec: pipeline_analytic}, in cases of polarization cross-spectra, we use the geometric mean of the temperature and polarization transfer function.

We reiterate our finding that the fits are nearly independent of the mask or power spectrum used in the simulations. This is why we can use a ``representative" mask and power spectra for all of DR6, and why the simulations ran here add negligible computational cost compared to the full simulation ensemble in \S\ref{sec: pipeline_sims}. Nevertheless, the procedure is ad-hoc and could be improved in the future.

\subsection{Parameters for Fiducial Signal Spectra} \label{apx: covmat_signal}

\begin{table}
    \centering
    \begin{tabular}{l|l|l}
        Parameter & Description & Value \\
        \hline
        \hline
        $log(10^{10}A_s)$ & Amplitude of primordial matter power spectrum & 3.044 \\
        $n_s$ & Power-law index of primordial matter power spectrum & 0.9649 \\
        $100\theta_{MC}$ & Acoustic scale & 1.04085 \\
        $\Omega_bh^2$ & Physical baryon density & 0.02237 \\
        $\Omega_ch^2$ & Physical cold dark matter density & 0.12 \\
        $\tau$ & Optical depth to reionization & 0.0544 \\
        \hline
        $A_{tSZ}$ & Amplitude of tSZ power spectrum & 2.971 \\
        $A_{kSZ}$ & Amplitude of kSZ power spectrum & 1.6 \\
        $A_p$ & Amplitude of CIB Poisson power spectrum & 7.614 \\
        $\beta_p$ & Spectral index of CIB Poisson power spectrum & 2.2 \\
        $A_c$ & Amplitude of CIB clustered power spectrum & 2.755 \\
        $\beta_c$ & Spectral index of CIB clustered power spectrum & 2.2 \\
        $T_c$ & Temperature of CIB modified blackbody & 9.6\,K \\
        $\xi$ & tSZ and CIB clustered correlation & 0.1 \\
        $A_s$ & Amplitude of point source power spectrum & 3.700 \\
        $\beta_s$ & Spectral index of point sources & -2.5 \\
        $A_d^{TT}$ & Amplitude of dust in $TT$ power spectrum & 8.83 \\
        $\alpha_d^{TT}$ & Power-law index of dust in $TT$ power spectrum & -0.6 \\
        $A_d^{TE}$ & Amplitude of dust in $TE$ power spectrum & 0.43 \\
        $\alpha_d^{TE}$ & Power-law index of dust in $TE$ power spectrum & -0.4 \\
        $A_d^{EE}$ & Amplitude of dust in $EE$ power spectrum & 0.165 \\
        $\alpha_d^{EE}$ & Power-law index of dust in $EE$ power spectrum & -0.4 \\
        $A_d^{TB}$ & Amplitude of dust in $TB$ power spectrum & 0.012 \\
        $\alpha_d^{TB}$ & Power-law index of dust in $TB$ power spectrum & -0.4 \\
        $A_d^{BB}$ & Amplitude of dust in $BB$ power spectrum & 0.116 \\
        $\alpha_d^{BB}$ & Power-law index of dust in $BB$ power spectrum & -0.4 \\
        $\beta_d$ & Spectral index of dust power spectrum & 1.5 \\
        $T_d$ & Temperature of dust modified blackbody & 19.6\,K \\
    \end{tabular}
    \caption{Parameters for the cosmology and likelihood foreground model of \citet{C20}, which is specified in terms of $D_{\ell}$. The six cosmological parameters above the solid line have a flat frequency dependence. The foreground parameters below the solid line correspond to a model normalized at 150\,GHz and $\ell=3,000$ for each component, and are rounded to three decimal places. There are no nuisance parameters in our model.}
    \label{tab: signal_model}
\end{table}

The fiducial signal power spectra used in the analytic covariance matrix (\S\ref{sec: pipeline_analytic_signal}) and the simulations (\S\ref{sec: pipeline_sims}) follow the cosmology and likelihood foreground model of \citet{C20} with model parameters given in Table \ref{tab: signal_model}. Unlike \citet{C20}, we convert the frequency-dependent model components into power spectra for each frequency band in DR6 by integrating over the full array passband rather than using effective frequencies.

\section{Updates for \texttt{mnms} Noise Simulations} \label{apx: mnms}

In this section, we state changes to the noise model implementations of \citet{mnms} that were necessary for the dr6.02 maps. A more complete account of the dr6.02 noise properties and models will be made available with the release of the maps.

In short, there are three main changes with respect to \citet{mnms}. Firstly, because of a change to the map pixel window in dr6.02 \citep[see][]{Naess2022}, the noise at the edge of the map footprint increased compared to dr6.01. Thus, in addition to masking observed pixels and pixels with near-zero crosslinking, we also mask the 10 arcmin bordering the footprint edge, and do not include them in noise models or simulations. Secondly, as is evident in Equation \ref{eq: noise_def}, the directional wavelet noise model includes the per-pixel noise standard deviations $\bm\sigma_i$, whereas these were excluded in \citet{mnms} for this model. Lastly, we found the PA6 inverse-variance maps to contain more arcminute-scale structure compared to PA4 or PA5. When used in either the tiled or directional wavelet noise models, these inverse-variance maps induced broadband mode-coupling that led to a percent-level but coherent noise power deficit across nearly all angular scales. We found that smoothing the PA6 inverse-variance maps with a two-arcminute Gaussian kernel eliminated the mode-coupling with no measurable unwanted side effects.

\subsection{Excess Large-Scale Polarization Noise Power in Simulations} \label{apx: mnms_Nl_excess}

\begin{figure}
    \centering
    \includegraphics[width=0.5\columnwidth]{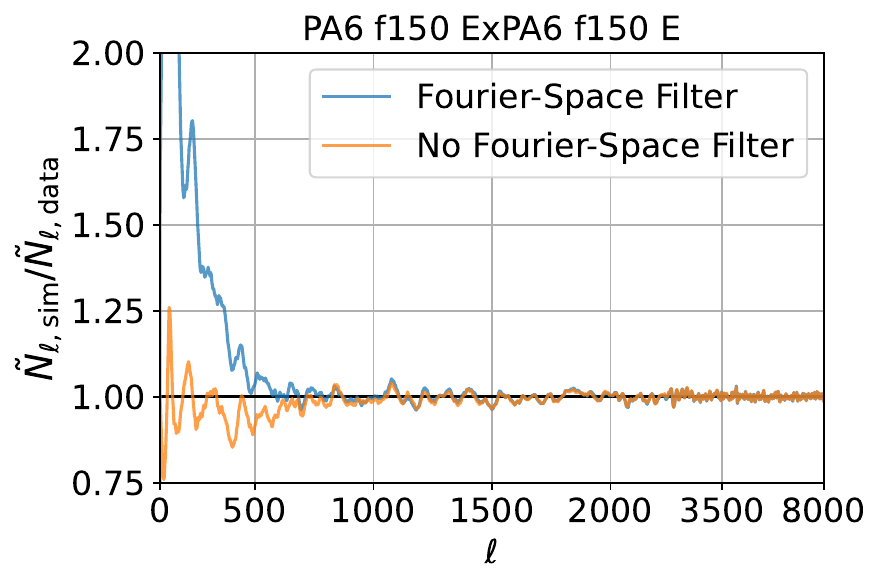}
    \caption{Ratio of polarization noise pseudospectra between simulations and data for PA6 f150. In the case of no Fourier-space filter being applied to the simulations or data, the ratio is consistent with unity to degree scales. Application of the Fourier-space filter to the simulations and data induces a large noise power excess in the simulations for scales larger than $\ell\approx500$.}
    \label{fig: res_kfilt_vs_no_kfilt}
\end{figure}

The origin of the excess large scale noise power in the simulations is due to the strong noise anisotropy pattern and the sharpness of the Fourier-space filter. To see how, we start by noting that in the absence of the Fourier-space filter, the simulated noise spectra agree much more closely with the data spectra, as is apparent in Figure \ref{fig: res_kfilt_vs_no_kfilt}. The noise power is greatest in the poorly cross-linked regions of the ACT scan strategy. As shown in Figure \ref{fig: form_pa5_f090_set0_2d_T_sims}, in these regions, the noise power is especially concentrated along the horizontal axis of 2D Fourier space. The Fourier-filter intentionally masks these modes: this more-optimally weights the data by cutting out a large fraction of the noise power but only a small fraction of the signal. However, this feature of the spatially-varying noise anisotropy is difficult to exactly reproduce in the simulations. If the simulations do not recover the exact noise anisotropy pattern, then after applying the filter, more noise power may persist in the unmasked modes in the simulations than in the data. In other words, application of the Fourier-space filter in this setting induces a mismatch in the per-$\ell$ average noise power between the simulations and the data where there was none before. Indeed, we find this to be the case for both noise models; Figure \ref{fig: res_kfilt_vs_no_kfilt} shows the result for the tiled simulations entering the baseline Monte Carlo covariance. In future work, we will optimize the noise modeling and the Fourier-space filter to enable accurate Monte Carlo covariance matrices to larger scales than those used in DR6. 

\section{Effect of Smoothing Diagonals of $\beta\neq\beta'$ Blocks of $\Sigma_R$} \label{apx: smooth_off_diags}

\begin{figure*}
    \centering
    \includegraphics[width=0.85\textwidth]{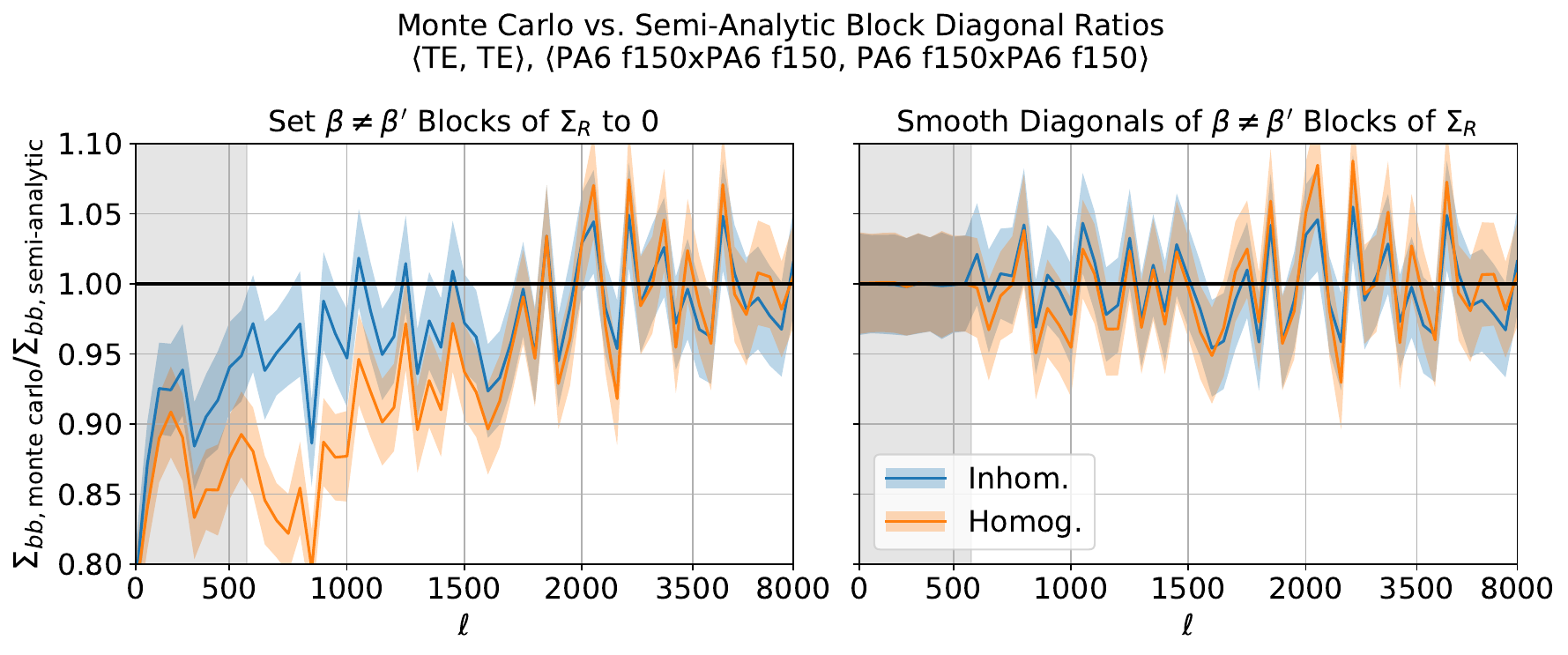}
    \caption{Ratios between the bin-wise diagonal of the Monte Carlo covariance matrix and the inhomogeneous and homogeneous semi-analytic covariances matrices, for the PA6 f150 x PA6 f150 $TE$ block-diagonal. \textit{Left:} In the case that the simulation-based correction only smooths the main diagonal of $\bm\Sigma_R$. \textit{Right:} In the nominal case, where all the bin-wise diagonals of $\bm\Sigma_R$ are smoothed, including the off-diagonal ($\beta'\neq\beta$) blocks.}
    \label{fig: mc_vs_semiana_block_diags}
\end{figure*}

We demonstrate the importance of including the percent-level $\bm\Sigma_R$ off-diagonals in the simulation-based correction of \S\ref{sec: pipeline_sim_correction}. Specifically, we examine the bin-wise diagonals ($b'=b$) of its off-diagonal blocks ($\beta'\neq\beta$). The primary effect of neglecting these small correlations --- by erroneously setting them to zero --- is shown in Figure \ref{fig: mc_vs_semiana_block_diags}: the amplitudes of the semi-analytic covariance bin-wise diagonals are too large compared to the Monte Carlo covariance. Notably, this has only a minor effect on the $d_{sim}^2$ distributions of Figure \ref{fig: res_chi2_red_ana_old}: the distributions have means of $1757.8 \pm 2.3$ and $1755.7 \pm 2.3$ for the inhomogeneous and homogeneous semi-analytic matrices, respectively. Both distribution means are within $0.5\%$ of the theoretical value of 1763. Nevertheless, large biases in the covariance matrix elements themselves can be problematic for subsequent analyses that manipulate the data vector and covariance, for example, by coadding them. Fortunately, Figure \ref{fig: mc_vs_semiana_block_diags} demonstrates that accounting for these small off-diagonals in our simulation-based correction results in good agreement between the Monte Carlo and semi-analytic covariance matrices for both the inhomogeneous and homogeneous analytic prescriptions.

\section{Effect of Point-Source Holes on the Covariance Matrix} \label{apx: pt_src_holes}

Although the results of \S\ref{sec: results} suggest that the NKA does not play a significant role in the ACT DR6 covariance, it is interesting to consider its effect regardless. For instance, the literature has long recognized the questionable validity of the NKA \citep[e.g.,][]{Efstathiou2004, Challinor2004, Brown2005}, especially due to analysis mask point-source holes \citep[e.g.,][]{Planck_XI_2015, Camphuis2022}. To measure the effect of the point-source holes, we compare an analytic covariance matrix built using analysis masks that lack the holes to a matrix using the nominal masks (see Figure \ref{fig: data_masks}). We do not apply the Fourier-space filter to the simulations or the data, and thus include no filter-correction transfer functions. This avoids two complications: as discussed in \S\ref{sec: pipeline_analytic_fourier}, the filter-correction can inadvertently absorb biases due to the NKA, and the noise simulations do not exhibit a large power excess at low-$\ell$. Compared to the filtered data, however, this does result in an overall steepening of the polarization noise power spectrum at low-$\ell$. Thus, this comparison provides an upper-bound on the impact of the NKA on the DR6 covariance. 

\begin{figure}[h]
    \centering
    \includegraphics[width=0.5\columnwidth]{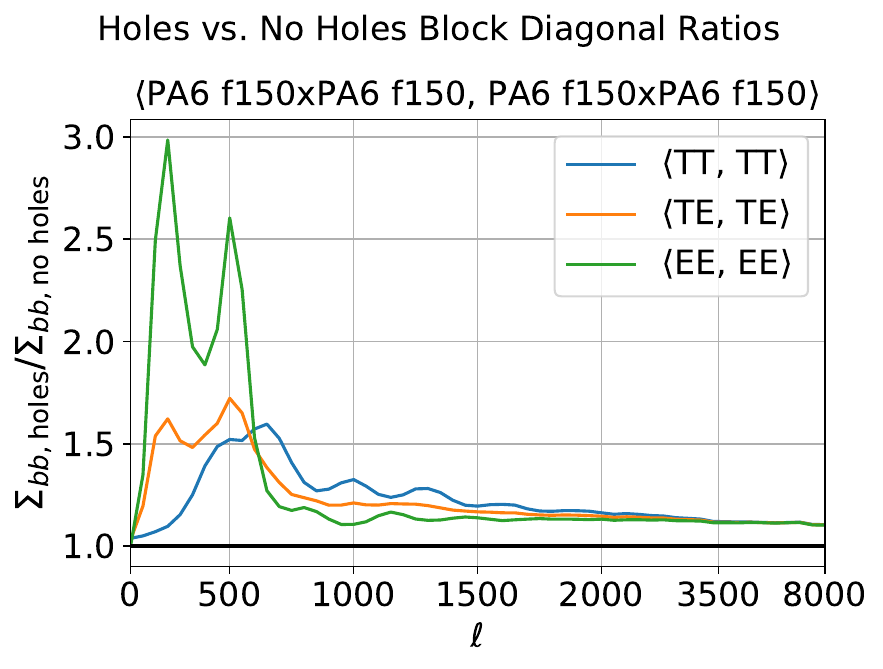}
    \caption{Ratios between analytic noise covariance matrix diagonals for an analysis mask including point-source holes to an analysis mask without point-source holes. For each polarization combination, the power spectrum variance increases when point-source holes are added to the mask, as expected.}
    \label{fig: disc_pt_src_holes_ana_vs_ana_ratio}
\end{figure}

\begin{figure*}
    \centering
    \includegraphics[width=0.85\textwidth]{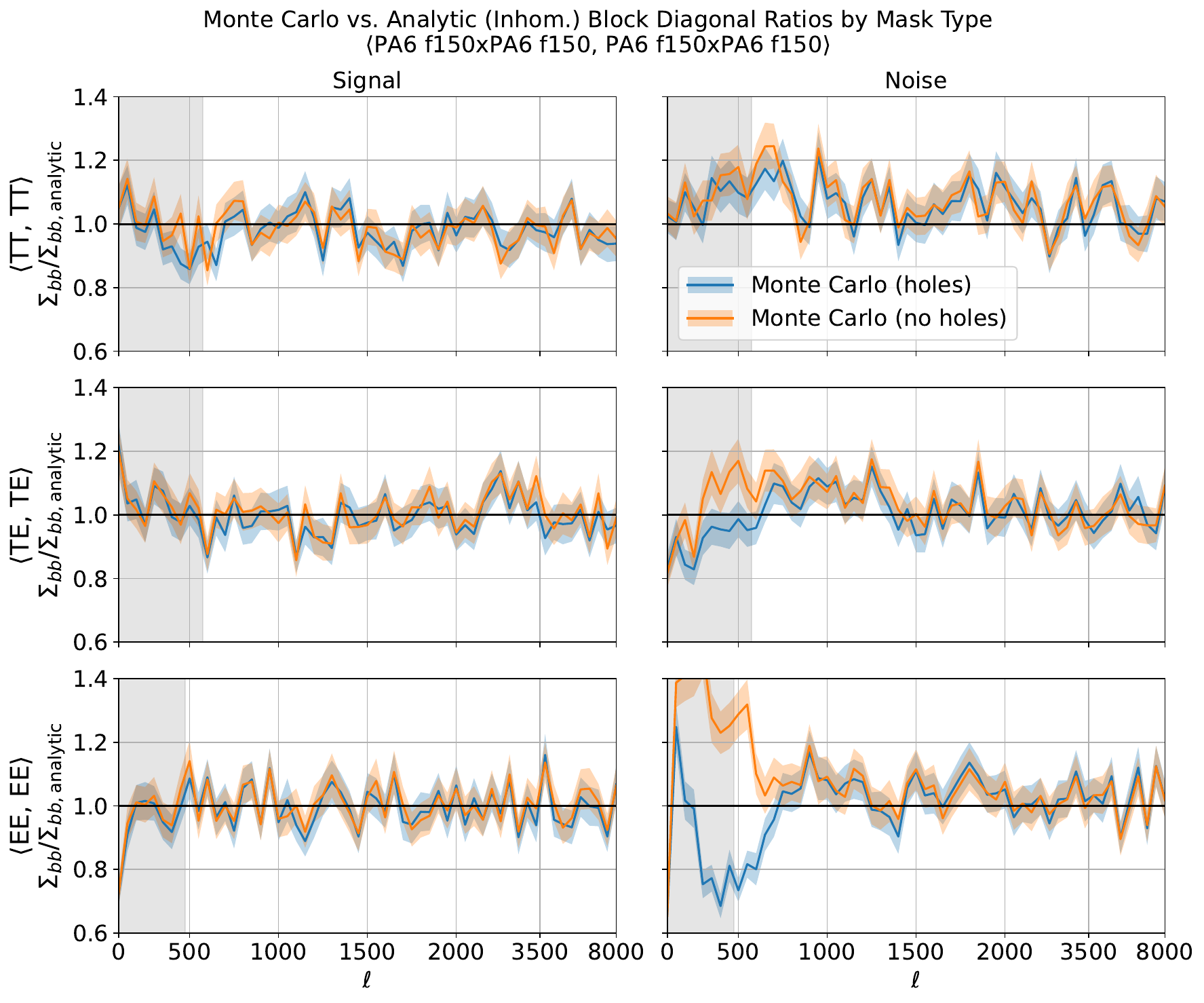}
    \caption{Ratios of Monte Carlo covariance matrix diagonals to analytic covariance matrix diagonals in the cases of an analysis mask with and without point-source holes. The format is analogous to Figure \ref{fig: res_mc_vs_ana_block_diags}. All results are for PA6 f150. With the exception of large-scale polarization covariance blocks, the addition of point-source holes to the analysis mask does not significantly downgrade the performance of the analytic covariance matrix.}
    \label{fig: disc_pt_src_holes_mc_vs_ana_ratios}
\end{figure*}

We first confirm our basic intuition that, due to the broadening of the mask power spectrum, an analysis including point-source holes increases the overall covariance magnitude in Figure \ref{fig: disc_pt_src_holes_ana_vs_ana_ratio}. The increase is most prominent in large-scale polarization. We then assess how the performance of uncorrected analytic covariance matrices, as measured against corresponding simulations, changes due to point-source holes in Figure \ref{fig: disc_pt_src_holes_mc_vs_ana_ratios}. In each case --- with and without point-source holes --- we construct a Monte Carlo covariance from the same 600 tile-based simulations. Thus, as is evident in Figure \ref{fig: disc_pt_src_holes_mc_vs_ana_ratios}, both cases share the same statistical fluctuations, further facilitating a direct comparison. For the signal part of the covariance, the point-source holes induce effectively no change in the performance of the analytic matrix: the holes and no-holes cases are indistinguishable, and both exhibit percent-level agreement between simulations and the analytic matrix. As discussed in \S\ref{sec: res_ana_validation}, this contrasts with \textit{Planck} \citep{Planck_XI_2015, Li2023}; however, this is likely due to a combination of the ACT bin-width, the smaller angular scales probed, and use of the INKA for the signal rather than the traditional NKA. 

The point-source holes have their largest effect on analytic matrix performance in the noise part of the covariance, especially large-scale polarization. We see the Monte Carlo covariance switches from being greater than, to being less than, the analytic covariance after adding point-source holes to the analysis mask. Neither effect is obvious in the right column of Figure \ref{fig: res_mc_vs_ana_block_diags}; the only difference with respect to the ``holes" case here being the averaging of signal and noise, and the Fourier-filter. Thus, it is likely that a combination of the shallower polarization noise power spectra, as well as the Fourier-filter correction's ability to absorb NKA-related biases, is operative in the baseline covariance matrix.

Interestingly, for large-scale polarization, even the no-holes case exhibits $\sim30\%$-level discrepancies between the Monte Carlo and analytic matrix, in spite of its smooth analysis mask. Coupled with the fact that, whether or not we apply the Fourier-space filter, the polarization noise power spectra are shallower than either temperature noise or signal power spectra suggests that the approximate noise model of Equation \ref{eq: approx_noise_def} is breaking down in this regime in addition to the NKA. We note that the noise anisotropy pattern in polarization is at least as strong as temperature in the poorly-crosslinked region. As we have shown in the case of the Fourier-filter, sharp anisotropies in 2D Fourier space can induce non-trivial effects at the covariance-level, regardless of their source (intrinsic to the noise, or introduced in the map processing). Further investigation is required to discern to what extent this effect could be present in the baseline covariance --- that is, when including the Fourier-filter. The practical result of this study is that since future large-aperture surveys, such as SO, will contend with similar noise properties as ACT, either more work is needed to better analytically account for the effect of stripy noise properties at the covariance level, or analyses will remain reliant on Monte Carlo covariance matrices for large-scale polarization.

\end{document}

%% file: authors.tex
\correspondingauthor{Zachary~Atkins}
\email{zatkins@princeton.edu}

\author[0000-0002-2287-1603]{Zachary~Atkins}
\affiliation{Joseph Henry Laboratories of Physics, Jadwin Hall, Princeton University, Princeton, NJ, USA 08544}

\author[0000-0002-0309-9750]{Zack~Li}
\affiliation{Lawrence Berkeley National Laboratory, 1 Cyclotron Road, Berkeley, CA 94720, USA}
\affiliation{Berkeley Center for Cosmological Physics, University of California, Berkeley, CA 94720, USA}
\affiliation{Canadian Institute for Theoretical Astrophysics, University of Toronto, McLennan Labs, Toronto, ON, M5S 3H8, Canada}

\author[0000-0002-4598-9719]{David~Alonso}
\affiliation{Department of Physics, University of Oxford, Denys Wilkinson Building, Keble Road, Oxford OX1 3RH, United Kingdom}

\author[0000-0003-2358-9949]{J.~Richard~Bond}
\affiliation{Canadian Institute for Theoretical Astrophysics, University of Toronto, McLennan Labs, Toronto, ON, M5S 3H8, Canada}

\author[0000-0003-0837-0068]{Erminia~Calabrese}
\affiliation{School of Physics and Astronomy, Cardiff University, The Parade, Cardiff, Wales CF24 3AA, UK}

\author[0000-0003-2856-2382]{Adriaan~J.~Duivenvoorden}
\affiliation{Max-Planck-Institut f\"{u}r Astrophysik, Karl-Schwarzschild-Str. 1, 85748 Garching, Germany}
\affiliation{Joseph Henry Laboratories of Physics, Jadwin Hall, Princeton University, Princeton, NJ, USA 08544}
\affiliation{Center for Computational Astrophysics, Flatiron Institute, New York, NY 10010, USA}

\author[0000-0002-7450-2586]{Jo~Dunkley}
\affiliation{Joseph Henry Laboratories of Physics, Jadwin Hall, Princeton University, Princeton, NJ, USA 08544}
\affiliation{Department of Astrophysical Sciences, Peyton Hall, Princeton University, Princeton, NJ, USA 08544}

\author[0000-0002-8340-3715]{Serena~Giardiello}
\affiliation{School of Physics and Astronomy, Cardiff University, The Parade, Cardiff, Wales CF24 3AA, UK}

\author[0000-0002-4765-3426]{Carlos Herv\'ias-Caimapo}
\affiliation{Instituto de Astrof\'isica and Centro de Astro-Ingenier\'ia, Facultad de F\'isica, Pontificia Universidad Cat\'olica de Chile, Av. Vicu\~na Mackenna 4860, 7820436 Macul, Santiago, Chile}

\author[0000-0002-9539-0835]{J.~Colin Hill}
\affiliation{Department of Physics, Columbia University, New York, NY 10027, USA}

\author[0000-0002-9429-0015]{Hidde~T.~Jense}
\affiliation{School of Physics and Astronomy, Cardiff University, The Parade, Cardiff, Wales CF24 3AA, UK}

\author[0000-0002-0935-3270]{Joshua~Kim}
\affiliation{Department of Physics and Astronomy, University of Pennsylvania, 209 South 33rd Street, Philadelphia, PA 19104, USA}

\author[0000-0001-7125-3580]{Michael~D.~Niemack}
\affiliation{Department of Physics, Cornell University, Ithaca, NY 14853, USA}
\affiliation{Department of Astronomy, Cornell University, Ithaca, NY 14853, USA}

\author[0000-0002-9828-3525]{Lyman~Page}
\affiliation{Joseph Henry Laboratories of Physics, Jadwin Hall, Princeton University, Princeton, NJ, USA 08544}

\author[0000-0002-2613-2445]{Adrien~La~Posta}
\affiliation{Department of Physics, University of Oxford, Denys Wilkinson Building, Keble Road, Oxford OX1 3RH, United Kingdom}

\author[0000-0002-6849-4217]{Thibaut~Louis}
\affiliation{Université Paris-Saclay, CNRS/IN2P3, IJCLab, 91405 Orsay, France}

\author[0000-0001-6606-7142]{Kavilan~Moodley}
\affiliation{Astrophysics Research Centre, University of KwaZulu-Natal, Westville Campus, Durban 4041, South Africa}
\affiliation{School of Mathematics, Statistics and Computer Science, University of KwaZulu-Natal, Westville Campus, Durban 4041, South Africa}

\author[0000-0002-5564-997X]{Thomas~W.~Morris}
\affiliation{Department of Physics, Yale University, New Haven, CT 06511, USA}
\affiliation{National Synchrotron Light Source II, Brookhaven National Laboratory, Upton, NY 11973, USA}

\author[0000-0002-4478-7111]{Sigurd~Naess}
\affiliation{Institute of Theoretical Astrophysics, University of Oslo, Norway}

\author[0000-0002-8149-1352]{Crist\'obal Sif\'on}
\affiliation{Instituto de F\'isica, Pontificia Universidad Cat\'olica de Valpara\'iso, Casilla 4059, Valpara\'iso, Chile}

\author[0000-0002-7567-4451]{Edward~J.~Wollack}
\affiliation{NASA Goddard Space Flight Center, 8800 Greenbelt Rd, Greenbelt, MD 20771 USA}